\pdfoutput=1
\newif\ifconfver
\confvertrue       
\ifconfver
\documentclass[10pt,twocolumn,twoside]{IEEEtran}
\else
\documentclass[11pt,draftcls,onecolumn]{IEEEtran}
\fi
\usepackage{algorithm2e}
\usepackage{makecell}
\usepackage{color,graphicx}
\usepackage{float}
\usepackage{multirow,amsmath,epsfig,amsfonts,amssymb,psfig,graphics,psfrag,theorem,calc,url,bm,cite}
\usepackage{stfloats,hyperref}
\usepackage{algorithmic}
\usepackage{diagbox} 
\usepackage{hhline}
\usepackage{subfigure}
\usepackage{booktabs}
\usepackage{multirow}
\usepackage{stmaryrd}
\graphicspath{{figure/}{figure/authors/}{figure/pseudo-color_images/}{figure/results/}}
\usepackage{mathtools}

\makeatletter
\def\multilimits@{\bgroup
\Let@
\restore@math@cr
\default@tag
\baselineskip\fontdimen10 \scriptfont\tw@
\advance\baselineskip\fontdimen12 \scriptfont\tw@
\lineskip\thr@@\fontdimen8 \scriptfont\thr@@
\lineskiplimit\lineskip
\vbox\bgroup\ialign\bgroup\hfil$\m@th\scriptstyle{##}$\hfil\crcr}
\def\Sb{_\multilimits@}
\def\endSb{\crcr\egroup\egroup\egroup}
\makeatother

{\end{list}}

\newlength{\twidth}
\ifconfver
\else
\fi
\newcommand{\ubf}[1]{\underline{\textbf{#1}}}

\definecolor{orange}{RGB}{255,107,0}

\newtheorem{Property}{Property}

\theorembodyfont{\rmfamily}

{\begin{list}{}{
\settowidth{\labelwidth}{\mbox{\textnormal{#1}}}%
\setlength{\leftmargin}{\labelwidth+\labelsep}}}%
{\end{list}}

\newcommand\bC{\ensuremath{{\bm C}}}

\newcommand\bU{\ensuremath{{\bm U}}}

\newcommand\bX{\ensuremath{{\bm X}}}
\newcommand\bY{\ensuremath{{\bm Y}}}

\newcommand\bu{\ensuremath{{\bm u}}}
\newcommand\bv{\ensuremath{{\bm v}}}

\newcommand\bx{\ensuremath{{\bm x}}}
\newcommand\by{\ensuremath{{\bm y}}}

\definecolor{orange}{RGB}{255,107,0}

\ifconfver

\author{Chia-Hsiang Lin,~\IEEEmembership{Senior Member,~IEEE}, Shih-Min Hsu,~\IEEEmembership{Graduate Student Member,~IEEE},\\Ching-Yun Liang,~\IEEEmembership{Graduate Student Member,~IEEE}, Jocelyn Chanussot,~\IEEEmembership{Fellow,~IEEE}, \\and Jhih-Yan Chen~\IEEEmembership{Graduate Student Member,~IEEE}}

\title{Hyperspectral Calibration Detection: A Novel Concept For Change Detection With Unsupervised Incremental Safe Pseudo-Labeling Implementation \vspace{-0.8cm}
\thanks{This study was supported by the Emerging Young Scholar Program (namely, the 2030 Cross-Generation Young Scholars Program) of National Science and Technology Council (NSTC), Taiwan, under Grant NSTC 115-2628-E-006-001.
We thank the National Center for Theoretical Sciences (NCTS) and the National Center for High-performance Computing (NCHC) for providing the computing resources.
\textit{(Corresponding author: Chia-Hsiang Lin.)}}
\thanks{C.-H. Lin is with the Department of Electrical Engineering, and with the Miin Wu School of Computing, National Cheng Kung University, Tainan 70101, Taiwan (R.O.C.) 
(e-mail: chiahsiang.steven.lin@gmail.com).}
\thanks{S.-M. Hsu, C.-Y. Liang, and J.-Y. Chen are with the Institute of Computer and Communication Engineering, Department of Electrical Engineering, National Cheng Kung University, Tainan, Taiwan (R.O.C.)
(e-mail: q38134015@gs.ncku.edu.tw; q36134182@gs.ncku.edu.tw; q36121113@gs.ncku.edu.tw).}
\thanks{J. Chanussot is with Inria, CNRS, Grenoble INP, LJK, Université Grenoble Alpes, 38000 Grenoble, France 
(e-mail:  jocelyn.chanussot@inria.fr).}
}

\else

\fi

\begin{document}

\bibliographystyle{IEEEtran}
\maketitle
\ifconfver \else \vspace{-0.5cm}\fi

\begin{abstract}
Hyperspectral change detection (HCD) has found numerous key applications, such as land cover monitoring.
The majority of benchmark HCD algorithms are semi-supervised methods, and some of them can even achieve very low sample labeling rates.
However, in some practical scenarios, such as those requiring immediate detection responses for onboard edge computing, we need to achieve the zero-label requirement as ground-truth labeling would not be available onboard for newly acquired images.
In this work, we propose a fully unsupervised HCD algorithm, together with a lightweight model, quite suitable for onboard detection missions.
Based on an iteratively augmented training set that safely collects some unchanged pixel samples, we learn an iteratively refined spectrum calibration function that eventually compensates the variability of acquisition conditions (often observed in bitemporal images), thereby making the changed pixels easily detectable by analyzing the calibrated spectra.
The proposed hyperspectral looping unsupervised calibration and incremental detection (HyperLUCID) algorithm is not only computationally efficient (around 1 to 2 orders of magnitude faster than most benchmark HCD methods), but has also achieved state-of-the-art results (around 93.6\% to 97.9\% overall accuracy) on several real benchmark HCD datasets.
Source codes: \url{https://github.com/IHCLab/HyperLUCID}.
\end{abstract}


\ifconfver \else \vspace{-0.0cm}\fi

\ifconfver \else \vspace{-0.5cm}\fi

\ifconfver \else  \fi

\vspace{-0.1cm}
\section{Introduction}\label{sec: introduction}
Hyperspectral remote sensing technologies have spurred numerous scientific and engineering applications \cite{wu2022uiu,8618436,8528557,wang2025generalized,chen2024flex,sun2021supervised}.
Hyperspectral images (HSIs), together with well-designed spectral analysis criteria, have truly outstanding material and object identifiability as validated through solid mathematical guarantees \cite{EMI,HISUN}.
Therefore, hyperspectral imagery is quite suitable for change detection and related missions.
Conventional hyperspectral change detection methods are often based on algebraic operations, image transformations, and classification techniques when identifying changed regions. 
These approaches often involve transforming the data into alternative representations or applying classification strategies to separate different classes \cite{deng2008pca,nielsen1998multivariate,nemmour2006multiple,huang2006extreme}.
However, these methods are often sensitive to threshold selection, rely on handcrafted feature representations, and depend heavily on the generalization ability of the chosen models, which may limit their robustness in complex real-world scenarios \cite{bruzzone2000automatic}.

\begin{figure}[t]
\begin{center}
\includegraphics[width=0.5\textwidth]{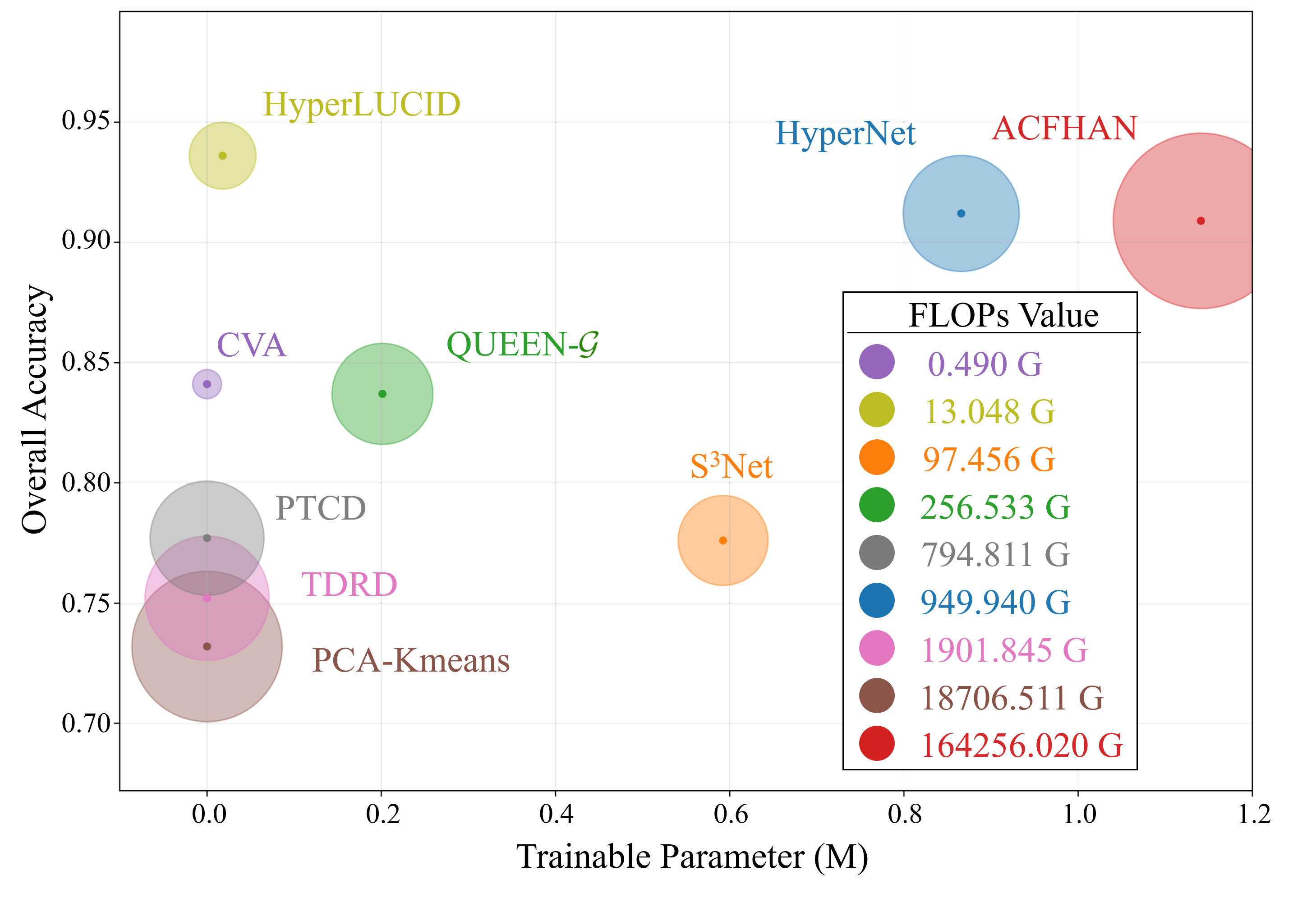}
\caption{Comparison of representative HCD methods on the large-scale Santa Barbara dataset in terms of overall accuracy (OA), trainable parameters, and FLOPs (i.e., the bubble radii).
Owing to the very wide range of the FLOPs, each bubble radius is presented in the log scale, with the corresponding exact raw value reported in the legend box.
We remark that there are frozen parameters (non-trainable) in the backbone of S$^3$Net, which are not counted, and those non-DL methods (i.e., CVA, PTCD, TDRD, and PCA-Kmeans) do not have trainable parameters.
Overall, the figure highlights the trade-off among accuracy, model size, and computational complexity, demonstrating the great superiority of the proposed HyperLUCID algorithm.}\label{fig:flopsradius}
\end{center}
\end{figure}

Recently, deep learning (DL) has become a dominant paradigm for hyperspectral change detection (HCD) due to its powerful representation learning capability \cite{guo2024saan,sun2021iterative,5427081,hu2023binary}. 
By learning hierarchical spatial-spectral features, DL-based methods can achieve strong performance in complex scenarios. 
Various architectures, including convolutional and attention-based models, have been proposed to further improve detection accuracy \cite{khelifi2020deep,lecun1989backpropagation,lin2019multispectral}. 
In general, the performance of these methods is strongly correlated with the amount of labeled training data, with larger datasets often leading to better accuracy \cite{SwinSUNet,TransUNetCD}.
This reliance on annotations significantly limits their applicability in real-world scenarios wherein labeled data are scarce or unavailable.
Nevertheless, obtaining accurately labeled datasets often poses great challenges. 
For instance, in some practical scenarios such as onboard edge computing, ground-truth labeling may not be available for newly acquired images. 
Similarly, in applications such as overgrazing detection, reliable annotations are often unavailable due to the lack of field investigation data. 
In these scenarios, it is necessary to develop approaches that do not rely on extensive labeled data. 
To this end, unsupervised, self-supervised, and semi-supervised methods have been proposed to mitigate the reliance on annotations by leveraging intrinsic data structures or limited supervision \cite{HUCD,SIAM,HyperNet,CODEHCD,QUEENG}.

In this work, we propose a fully unsupervised hyperspectral change detection (HCD) algorithm, together with a lightweight model (cf. Figure \ref{fig:flopsradius}), quite suitable for onboard edge-computing missions. 
We iteratively augment the training set that safely collects some unchanged pixel samples, and accordingly train an iteratively refined hyperspectral calibration function, aiming at compensating the variability of acquisition conditions often observed in bitemporal images.
We develop an unsupervised mechanism (cf. Algorithm \ref{algo:HyperLUCID}) to obtain the hyperspectral calibration function, thereby making the changed pixels easily detectable by analyzing the calibrated spectra.
The concept, which to the best of our knowledge has not been explored in prior HCD literature, is established in Section \ref{sec:problem}, and realized with an unsupervised implementation in Section \ref{sec:algo}.
The induced hyperspectral looping unsupervised calibration and incremental detection (HyperLUCID) algorithm is computationally efficient, and has achieved state-of-the-art detection results on several real benchmark HCD datasets, as experimentally demonstrated in Section \ref{sec: experiment}.
Finally, concluding remarks are drawn in Section \ref{sec: conclusion}.

\subsection{Related Work}\label{relatedwork}

In this section, we report HCD methods ranging from those conventional techniques to unsupervised DL methods.
% CVA
For example, changed vector analysis (CVA) \cite{bovolo2006theoretical} is one of the most well-known algebra-based approaches. 
It analyzes the difference vectors between corresponding pixels in bitemporal images to generate the change map. 
Moreover, it introduces a polar domain formulation, and theoretically analyzes the statistical distributions of changed and unchanged pixels under this representation. 
On the other hand, transformation-based methods convert bitemporal images into specific feature domains to highlight their variations. 
Commonly used techniques include principal component analysis (PCA) \cite{deng2008pca} and multivariate alteration detection (MAD) \cite{nielsen1998multivariate}, where the former reduces dimensionality by projecting data onto orthogonal components retaining most signal energies, and the latter maximizes the difference between unchanged and changed areas through canonical correlation analysis.
% PCA Kmeans
Building upon this idea, PCA-Kmeans \cite{PCAKMeans} applies PCA to the difference image to extract the most informative features, followed by K-means clustering to partition the feature space into changed and unchanged groups.
Classification-based methods treat change detection as a classification problem, employing classifiers such as support vector machines (SVM) \cite{nemmour2006multiple} and extreme learning machine (ELM) \cite{huang2006extreme} to distinguish the changed and unchanged regions.

% CNN
Among the supervised DL methods, a general end-to-end two-dimensional convolutional network (GETNET) \cite{wang2018getnet} introduced a mixed-affinity matrix to effectively extract cross-channel gradient information, allowing efficient processing of multi-source data simultaneously.
Furthermore, several recent change detection methods are based on different attention mechanisms. 
Given the inherent symmetricity of bitemporal HSIs, a novel cross-temporal interaction symmetric attention network (CSANet) \cite{song2022csanet} has been proposed to learn joint spatial–spectral–temporal features. 
By employing a siamese two-dimensional CNN, the CSANet effectively enhances the feature discrimination ability of the changed features.
In addition to the cross-temporal interaction, the fusion of spatial and spectral features has also benefited the detection task.
For instance, the multilevel encoder–decoder attention network (ML-EDAN) \cite{qu2021multilevel} integrates hierarchical features across multilevel with spatial-spectral information to facilitate accurate change detections.
Another representative method is the spectral-spatial-attention siamese network (SSA-SiamNet), which leverages the convolutional block attention modules for identifying informative spectral channels and analyzing spatially informative regions \cite{SSA-SiamNet}.
In SwinSUNet, they combine swin transformer network with siamese U-shaped structure to learn global information, while TransUNetCD adopts a supervised encoding-decoding hybrid transformer architecture\cite{SwinSUNet,TransUNetCD}.
TransUNetCD not only enhances global contextual information through tokenized patch embeddings, but also captures feature discrimination with skip connections.

Though these supervised DL methods demonstrate strong capability in learning complex spatial-spectral representations, obtaining accurately labeled datasets often poses great challenges.
Thus, unsupervised methods are desired.
As a representative, TDRD \cite{TDRD} introduces a tensor-based approach for HCD.
After representing multitemporal HSIs as high-order tensors, it applies a third-order Tucker decomposition to extract high-order principal components (PCs). 
A singular value accumulation strategy is then used to select the dominant components, followed by spectral angle analysis to determine whether the bitemporal pixel pairs are changed or not. 
Similarly, PTCD \cite{PTCD} also employs tensor decomposition and reconstruction strategies, while adopting a patch-based approach to capture local spatial structures by exploiting the non-overlapping similarity among neighboring patches.
Another novel hyperspectral sparse unmixing and unsupervised deep clustering based change detection (HUDC-CD) framework \cite{HUCD} introduces the powerful unmixing theory \cite{SIAM} and uses non-local (NL) means as a spatial regularizer to enhance an unmixing-based detector. 
However, traditional NL means require exhaustive search across the entire image, which is computationally expensive.
To address this issue, the authors of \cite{HUCD} judiciously propose an unsupervised deep clustering approach that groups similar pixels into homogeneous regions, thereby constraining the NL-means searching space and improving computational efficiency.
Finally, change detection is performed by identifying altered endmembers using a non-dominated sorting genetic algorithm (NSGA).

Beyond fully unsupervised approaches, recent studies have explored learning paradigms that further leverage unlabeled data through self-supervision.
% Introduce self-supervised peer-methods
There is a special class of approaches that generate pseudo-labels directly from the data itself, allowing the model to learn without requiring manual annotations.
This class is commonly referred to as self-supervised learning.
For example, HyperNet \cite{HyperNet} utilizes self-supervised strategy specifically designed for pixel-level hyperspectral change detection.
Unlike conventional self-supervised approaches that focus on patch-level representations, HyperNet operates on full multitemporal HSIs and compares spatial-spectral features pixel by pixel. 
Moreover, it introduces a spatial–spectral attention module to independently capture spatial correlations and discriminative spectral characteristics.
To facilitate the training, a novel focal cosine loss is proposed to emphasize hard positive samples and improve training efficiency.
Moreover, superpixel-guided self-supervised network ($\text{S}^3$Net) \cite{S3Net} also adopts a self-supervised framework guided by superpixel segmentation for change detection.
It first constructs a pseudo-color image via principal component analysis of multitemporal inputs, and segments it into superpixels to define homogeneous spatial regions.
A siamese network with shared weights is then trained in a self-supervised manner to extract object-level multiscale spatial feature differences.
Finally, a weighted fusion of pixel-level and object-level differences produces the final change map.
Also, hypergraph-based method has been proposed to model multitemporal hyperspectral images by capturing high-order hyperspectral-spatial-temporal (HyperSST) correlations among pixels \cite{A}.
HyperSST adopts self-supervised learning strategies, including contrastive learning for spectral-spatial features and generative learning for temporal features, to learn discriminative representations without requiring labels.
The learned joint representations are then utilized for downstream change detection (CD), providing comprehensive modeling of both structural and temporal variations.

% Introduce semi-supervised peer-methods
While self-supervised methods rely solely on the structure of unlabeled data, semi-supervised strategies benefit from a small amount of annotated samples.
Among them, the CODE algorithm \cite{CODEHCD,CODE} integrates convex optimization (CO) and deep learning (DE), and combines this CODE theory with graph neural network (GNN), thereby achieving remarkable change detection results with very low sample labeling rates.
Furthermore, the QUEEN-$\mathcal{G}$ algorithm \cite{QUEENG} pioneers the integration of quantum deep networks (QUEEN) \cite{lin2026underdetermined,HyperKING} into GNN-based detection by combining unitary computing features from quantum models with graph-based spatial inferences.
It consists of a graph feature learning module operating at the superpixel level to capture spatial correlations, and a quantum feature learning module at the pixel level to extract complementary detailed textures.
A quantum classifier is then employed in QUEEN-$\mathcal{G}$ to achieve highly effective detection.
In addition, an unsupervised method termed Stripe-CD \cite{B} is proposed based on untrained networks and optimization modeling.
Stripe-CD employs a dual-branch untrained convolutional network to extract deep spectral-spatial difference features, and further constructs a stripe-based feature space in which change extraction is formulated as an optimization problem and solved via alternating direction method of multipliers (ADMM), thereby enabling unsupervised HCD.

\begin{figure}[t]
\begin{center}
\includegraphics[width=0.5\textwidth]{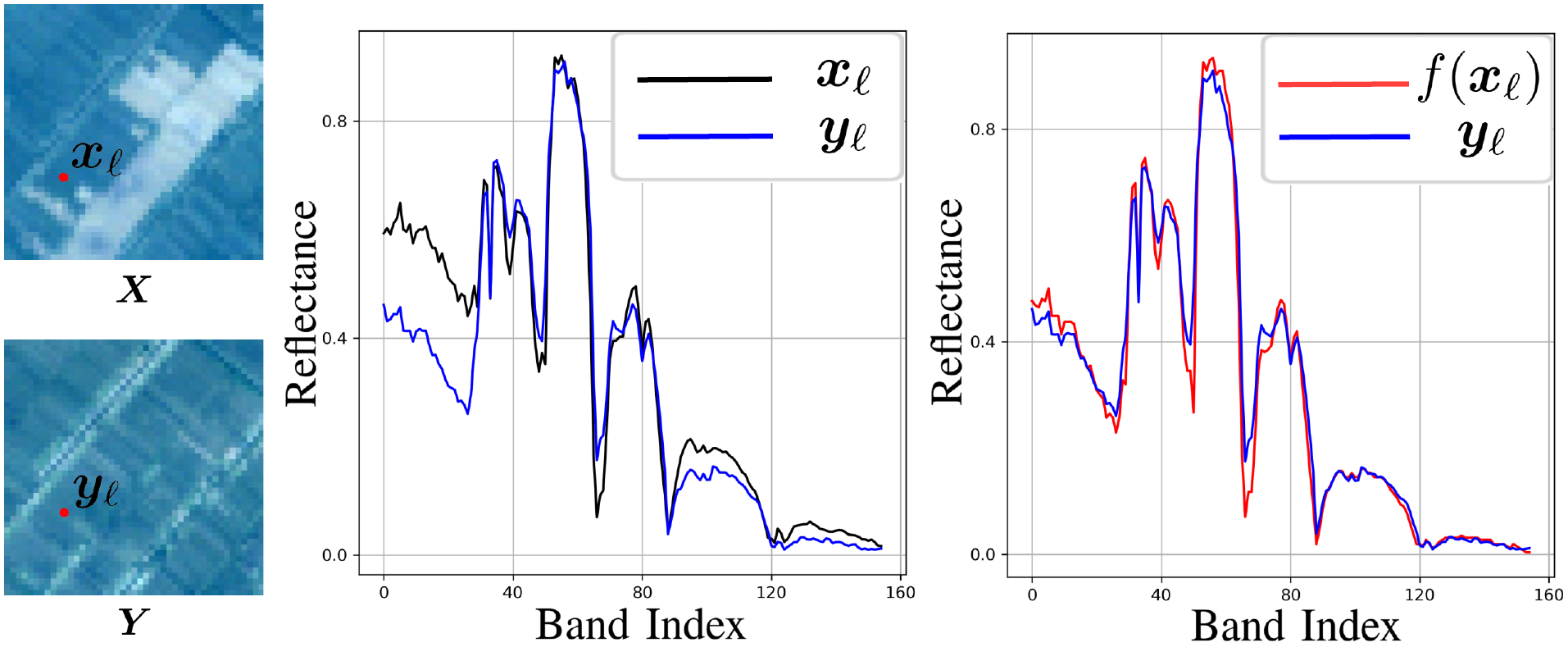}
%\vspace{0.05cm}
\caption{Graphical illustration of the HyperCAD concept for HCD.
Even for an unchanged location $\ell$, marked by the red dots in $(\bX,\bY)$, the corresponding pixels $(\bx_\ell,\by_\ell)$ look quite deviated with a larger distance $\textrm{dist}\left(\bx_\ell,\by_\ell\right)$, resulting in a misdetection of change.
If one can learn an effective hyperspectral calibration function $f$, the calibrated pixel $f(\bx_\ell)$ becomes not distinguishable from
$\by_\ell$, leading to a smaller distance $\textrm{dist}\left( f(\bx_\ell) ,\by_\ell\right)$, thereby successfully detecting $\ell$ as an unchanged location.
The questions then become how to obtain $f$ and how to achieve so unsupervisedly and efficiently.
}\label{fig:HyperCAD}
\end{center}
\end{figure}

\vspace{-0.1cm}
\section{The Proposed HyperLUCID Algorithm for Unsupervised HCD} \label{sec:method}

\subsection{Problem Description and Challenges}\label{sec:problem}

Given the HSI pair $(\bX,\bY)$ acquired at different time instances but over the same spatial scene, the hyperspectral change detection (HCD) problem aims at identifying those pixels that are changed between $\bX=[\bx_1,\dots,\bx_L]\in\mathbb{R}^{M\times L}$ and $\bY=[\by_1,\dots,\by_L]\in\mathbb{R}^{M\times L}$.
Here, $M$ is the number of hyperspectral bands, and $L$ is the number of pixels.
If $(\bX,\bY)$ are acquired under exactly the same condition, the HCD problem can be simply solved by comparing each pixel pair $(\bx_\ell,\by_\ell)$ for $1\leq\ell\leq L$, followed by simply discriminating the $\ell$th pixel as a changed one if, and only if, 
\begin{equation}\label{eq:DI}
\|\bx_\ell-\by_\ell\|_2 > \epsilon
\end{equation}
for some given threshold $\epsilon>0$ (i.e., $\bx_\ell$ and $\by_\ell$ are too distinctive) \cite{CDADMM}.
Thus, a most fundamental and straightforward way to solve the CD problem is to computationally align the acquisition conditions, which will be formulated below and achieved using elegantly simple mechanisms.

In real-world applications, however, there are numerous practical challenges, as $(\bX,\bY)$ are seldom acquired under similar conditions.
For example, assuming the location of the $\ell$th pixel is unchanged, the quantity $\|\bx_\ell-\by_\ell\|_2$ could still be large if the illumination condition when acquiring $\by_\ell$ is too deviated from that when acquiring $\bx_\ell$.
This naturally motivates us to find/learn a hyperspectral calibration function $f$, under which the calibrated pixel at the first time instance [i.e., $f(\bx_\ell)$] and the pixel acquired at the second time instance (i.e., $\by_\ell$) satisfy the two relations in Property \ref{prop:GoodMapping}.
\begin{Property}\label{prop:GoodMapping}
An ideal calibration function $f$ holds with the following two relations:
\begin{enumerate}
\item 
The calibrated pixel $f(\bx_\ell)$ is not distinguishable from $\by_\ell$, if the $\ell$th pixel is an unchanged one.

\vspace{0.1cm}

\item 
The calibrated pixel $f(\bx_\ell)$ is clearly distinguishable from $\by_\ell$, if the $\ell$th pixel is a changed one.
\hfill$\square$
\end{enumerate}
\end{Property}

\noindent
If such a function $f$ is available, one can adopt the above two properties to modify \eqref{eq:DI} for obtaining a more robust but still simple detection criterion.
Specifically, the HCD problem can be elegantly solved by discriminating the $\ell$th pixel as a changed one if, and only if, 
\begin{equation}\label{eq:DI-modify}
\textrm{dist}\left(f(\bx_\ell),\by_\ell\right) > \epsilon
\end{equation}
where $\textrm{dist}(\bu,\bv)$ is a distance function measuring the difference between $\bu$ and $\bv$.
See Figure \ref{fig:HyperCAD} for an illustration.
An ideal set of training samples for obtaining $f$ will be designed in Section \ref{sec:algo}.

The above concept of hyperspectral calibration detection (HyperCAD), illustrated in Figure \ref{fig:HyperCAD}, induces the following questions.
First, if one allows a semi-supervised setting as assumed in the majority of existing HCD works \cite{QUEENG}, $f$ can be learned from those partially labeled pixels.
However, to facilitate more practical applicability, can the above HyperCAD concept be implemented under a fully unsupervised manner?
Second, can the distance function ``$\textrm{dist}(\cdot,\cdot)$'' in \eqref{eq:DI-modify} be properly customized for HCD?
Third, can the threshold ``$\epsilon>0$'' be unified (i.e., data-independent), rather than manually tuned?

We will address all the above three questions in Section \ref{sec:algo}, thereby implementing the HyperCAD concept to achieve a fast, effective, and user-friendly algorithm (cf. Algorithm \ref{algo:HyperLUCID}) with state-of-the-art HCD performance.

\subsection{Unsupervised Implementation of HyperCAD Concept for HCD Algorithm Design}\label{sec:algo}

As discussed in Section \ref{sec:problem}, obtaining the hyperspectral calibration function $f$ is not difficult if we allow supervised or semi-supervised approaches, which, however, hampers the development of a user-friendly HCD method.
Note that collecting the ground-truth (GT) labels could be impractical or even infeasible in some scenarios.
For example, given the data $\bX$ acquired at the first time instance, the onboard detection would require an immediate HCD result right after the acquisition of $\bY$, for which we aim to return the HCD results in a few seconds, rather than waiting for hours or days to collect the GT labels for supervised or semi-supervised learning.

Inspired by the gradually-augmented sample selection strategy proposed in \cite{su2015active}, originally tested on the datasets of handwritten digits and stem cells, we will design a fully unsupervised HCD algorithm to implement the novel HyperCAD concept.\footnote{The sample selection criterion used in \cite{su2015active} induces an NP-hard combinatorial optimization problem with human interventions, while our method adopts the SAM-based criterion (cf. \eqref{eq: SAM}) in a fully automatic manner.
% polynomial-time computing
Our SAM-based criterion only requires a complexity linear to the number of samples, because it computes the SAM distance $L$ times for selecting the 20\% low-SAM pixels in $\bm\Omega^{t+1}$ (cf. Algorithm \ref{algo:HyperLUCID}).}
Specifically, our HCD algorithm will be based on a gradually-augmented index set $\bm\Omega\subseteq\{1,\dots,L\}$ of unlabeled pixels for unsupervised learning of the calibration function $f$.
We partition the $L$ pixels into changed ones $\bC$ and unchanged ones $\bU$, where $\bC\cup\bU=\{1,\dots,L\}$ and $\bC\cap\bU=\emptyset$.
Thus, an ideal index set $\bm\Omega^\star$, associated with the data pair $(\bX,\bY)$, is expected to satisfy the following conditions:
\begin{enumerate}

\vspace{0.15cm}

\item[(A)]
$\bm\Omega^\star$ contains only those unchanged pixels, i.e., $\bm\Omega^\star\subseteq\bU$.

\vspace{0.15cm}

\item[(B)]
$\bU\setminus\bm\Omega^\star$ should be as close to $\emptyset$ as possible.

\vspace{0.15cm}

\item[(C)]
$\bm\Omega^\star$ may be iteratively computed as $\bm\Omega^\star$ (and also $\bU$) is unknown in the unsupervised setting, implying that $\bm\Omega^\star$ can be represented as $\bm\Omega^\star:=\lim_{t\rightarrow\infty}\bm\Omega^t$ with $\bm\Omega^t$ being the estimate of $\bm\Omega^\star$ at the $t$th iteration.

\vspace{0.15cm}

\item[(D)]
$\bm\Omega^t$ has to be computed as fast as possible, thereby avoiding a slow estimation procedure for obtaining $\bm\Omega^\star$.
\end{enumerate}

Regarding the relation between $f$ and $\bm\Omega^\star$, note that to learn the calibration function $f$, we need to identify some unchanged pixels to serve as the learning samples, as illustrated in Figure \ref{fig:HyperCAD}.
The aim of the ideal index set $\bm\Omega^\star$ (and $\bm\Omega^t$) is to identify some safe unchanged samples in a computationally efficient manner, as defined by the four conditions above, whose motivations will be illustrated below.
Also, how the four conditions help efficiently obtain the ideal $f$ defined in Property \ref{prop:GoodMapping} will also be discussed below.

To see the motivation of condition (A), note that an ideal $f$ should satisfy Property \ref{prop:GoodMapping}.
Let us prove the necessity of (A) by contradiction.
If condition (A) is violated, then $f$ learned from $\bm\Omega^\star$ would map a changed pixel $\bx_\ell$ to a spectrum $f(\bx_\ell)$ that is not clearly distinguishable from $\by_\ell$, meaning that $f$ does not hold with Property \ref{prop:GoodMapping}.2)---a contradiction.
So, condition (A) is a necessary one.
To understand the motivation of condition (B), we also prove it by contradiction.
If condition (B) is violated, then there exist many unchanged pixels (in $\bU$) that are not collected in $\bm\Omega^\star$.
Thus, the $f$ learned from $\bm\Omega^\star$ would not have the capability of mapping most unchanged pixel $\bx_\ell$ (in $\bU$) to a spectrum $f(\bx_\ell)$ similar to $\by_\ell$, meaning that Property \ref{prop:GoodMapping}.1) does not hold true---a contradiction.
Therefore, condition (B) is a necessary one.
To see the motivation of condition (C), recall that we are pursuing an unsupervised setting.
So, $\bm\Omega^\star$ is unknown.
In such a case, a natural way is to iteratively estimate $\bm\Omega^\star$.
To facilitate the estimation efficiency, we hope that the iteration result could directly serve as the estimation result, i.e., $\bm\Omega^\star:=\lim_{t\rightarrow\infty}\bm\Omega^t$.
One simple way is to find an implementation trick such that $\{\bm\Omega^t\}_{t=1}^\infty$ forms an increasing sequence, i.e.,
\begin{equation}\label{eq:increasingOmega}
\bm\Omega^t\subseteq\bm\Omega^{t+1},
\end{equation}
implying that $\bm\Omega^\star$ can be conveniently estimated using the set union as $\bm\Omega^\star:=\cup_{t=1}^\infty\bm\Omega^t=\lim_{t\rightarrow\infty}\bm\Omega^t$ [i.e., condition (C)].
As for condition (D), it is surely a desired one to achieve fast implementation.
Therefore, conditions (A) and (B) are set to meet Property \ref{prop:GoodMapping} defining an ideal $f$, while conditions (C) and (D) are more implementation-oriented.

\begin{figure}[t]
\begin{center}
\includegraphics[width=0.48\textwidth]{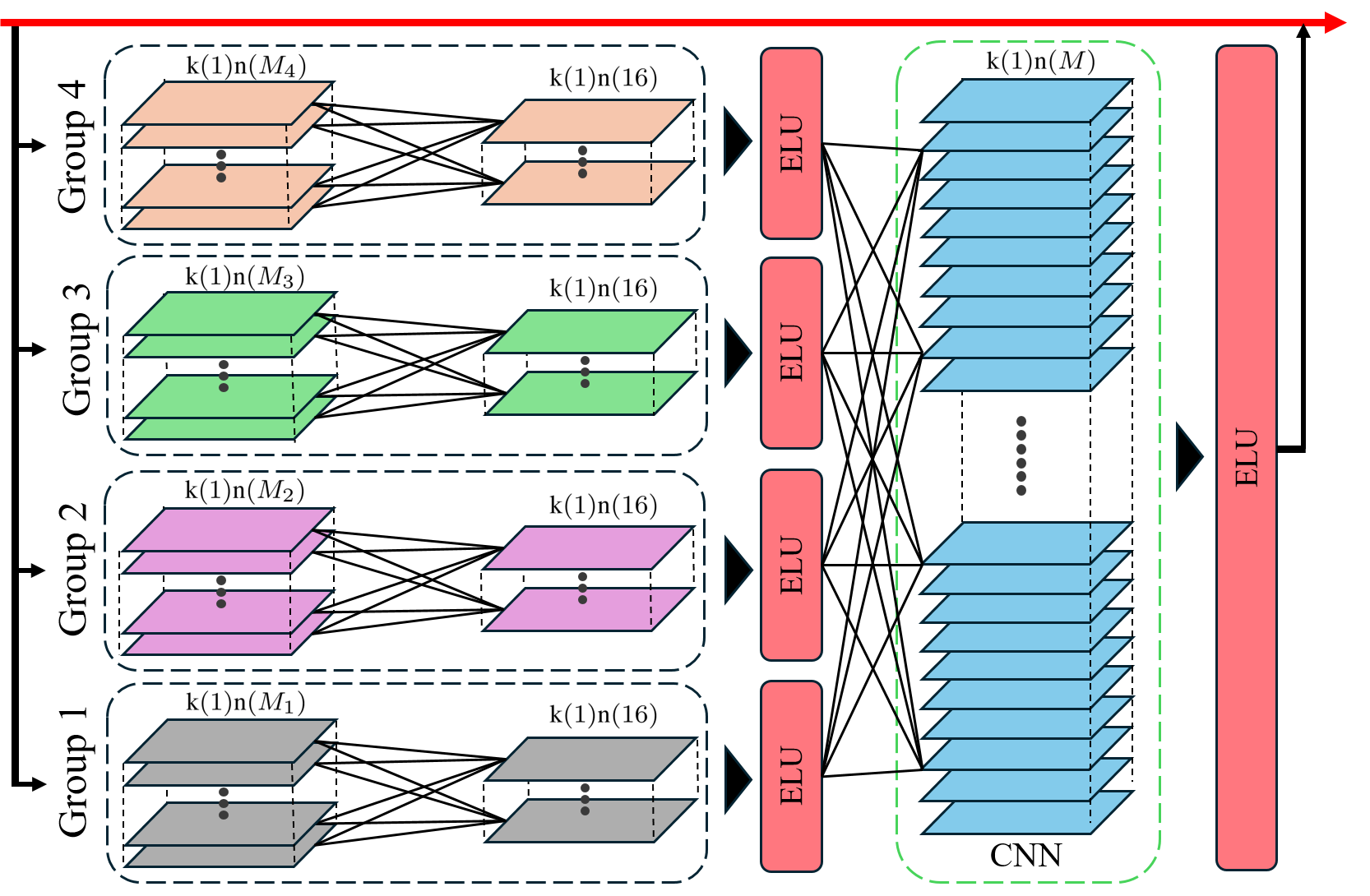}
\caption{A lightweight simple network architecture $f$ for ensuring fast computation of $\bm\Omega^t$ [cf. condition (D)].
The first phase (i.e., a two-layer encoder) employs the trick of group convolution that significantly reduces the number of parameters, thereby facilitating the computational efficiency, where $M_1=M_2=M_3\triangleq \lfloor \frac{M}{4} \rfloor$ and $M_4\triangleq M-3 M_1$ divide the $M$ input channels into four groups.
Following conventional usage \cite{CODE}, the notation ``$\textrm{k}(s)\textrm{n}(c)$'' denotes a convolutional module with $s \times s$ kernel size and $c$ output channels.
Exponential linear unit (ELU) is empirically found to be more effective for our task, and is hence selected as the activation function.
The second phase (i.e., a single-layer decoder) then integrates all the groups to allow the inter-group feature interaction, which also plays an essential role in the hyperspectral calibration task of $f$, as will be experimentally verified.
As $f$ endeavors to learn the calibration from $\bm\Omega^\star$ (which safely collects some unchanged samples $\ell$), its input $\bx_\ell$ and the calibrated result $f(\bx_\ell)$ should hold some resemblance for unchanged location (cf. the right plot of Figure \ref{fig:HyperCAD}).
Therefore, learning the residual ``$f(\bx_\ell)-\bx_\ell$'' would be more effective, motivating us to add a skip connection (i.e., the red arrow) over the two phases (to copy the information of $\bx_\ell$ directly to the output).
This simple residual strategy significantly accelerates the HyperLUCID algorithm, as will be proved in Section \ref{sec:ablation}.
The above concise implementation of $f$ will be experimentally demonstrated to be quite effective for HCD.}\label{fig:netf}
\end{center}
\end{figure}

Condition (D) requires that $f$ (for computing $\bm\Omega^t$) should be as ``simplified'' as possible.
In other words, if $f$ is implemented as a lightweight simple network, the condition (D) would be directly ensured.
Specifically, based on the empirically selected activation function and the trick of group convolution for lightweight implementation, the architecture of the hyperspectral calibration network $f$ is depicted in Figure \ref{fig:netf}.
As $f$ endeavors to learn the calibration from $\bm\Omega^\star$ [safely collecting some unchanged locations $\ell$; cf. condition (A)], its input $\bx_\ell$ and the returned calibration $f(\bx_\ell)$ should hold some resemblance (cf. Figure \ref{fig:HyperCAD}), motivating us to adopt the residual learning structure depicted in Figure \ref{fig:netf}.

After fulfilling condition (D) with the concise architecture, the remaining task is to design an iterative algorithmic scheme to meet conditions (A), (B), and (C).
The proposed algorithm, termed as hyperspectral looping unsupervised calibration and incremental detection (HyperLUCID), is summarized in Algorithm \ref{algo:HyperLUCID}, where the $(t+1)$th training stage in Line 6 adopts the outlier-robust L1-norm loss function defined as 
\begin{equation}\label{def:loss}
\frac{1}{|\bm\Omega^{t+1}|}\sum_{\ell\in\bm\Omega^{t+1}}\|f^{t+1}(\bx^t_\ell)-\by_\ell\|_1.
\end{equation}
This simple masked loss function not only ensures fast computation as required in condition (D), but also elegantly introduces robustness against the violation of condition (A) as will be discussed in Property \ref{prop:L1robust}.
Simply speaking, the proposed Algorithm \ref{algo:HyperLUCID} aims at iteratively identifying $f$ which can well calibrate the pixel $\bx_\ell$ into $\by_\ell$ for those unchanged locations $\ell$ specified by $\bm\Omega^{t+1}$, according to Property \ref{prop:GoodMapping}.

\begin{algorithm}[t]
\DontPrintSemicolon
\caption{The Proposed HyperLUCID Algorithm}\label{algo:HyperLUCID}
\begin{algorithmic}[1]  
\STATE \textbf{Given} bitemporal data $(\bX,\bY)$, and default hyperparameter settings, including the ratio of safe samples $R_\textrm{safe}:=20\%$, the threshold standard deviation $\eta_1:=0.05$, and the threshold SAM value $\eta_2:=0.2$.

\STATE 
Set $t\coloneqq0$, and denote $\bX^t:=[\bx_1^t,\dots,\bx_L^t]\in\mathbb{R}^{M\times L}$. 
Initialize $\bX^t := \bm X$ and $\bm\Omega^t:=\emptyset$ (the empty set).

\REPEAT 

\STATE Update $\bm\Omega^{t+1}$ as the set of pixel indices corresponding to the $R_\textrm{safe}:=20\%$ lowest SAM values between $(\bX^t,\bY)$.

\STATE Update the set union as $\bm\Omega^{t+1}:=\bm\Omega^{t+1}\cup\bm\Omega^{t}$ in order to ensure \eqref{eq:increasingOmega}.

\STATE Update the lightweight network $f^{t+1}$ (cf. Figure \ref{fig:netf}) using the training data pairs $(\bx^t_\ell,\by_\ell)$ with $\ell\in\bm\Omega^{t+1}$ based on adaptive moment estimation (Adam) \cite{kingma2014adam}.

\STATE Update $\bm X^{t+1} := f^{t+1}(\bX^t)$.

\STATE $t\coloneqq t+1$

\UNTIL the predefined stopping criterion is met (i.e., until the difference of standard deviation of the SAM values is less than $\eta_1$).

\STATE Compute the $\ell$th entry of $\bC^\star:=(c_1^\star,\dots,c_L^\star)\in\{0,1\}^L$ as $c_\ell^\star \triangleq \llbracket \textrm{SAM}(\bx_\ell^t,\by_\ell) > \eta_2 \rrbracket$, where $\llbracket \cdot \rrbracket$ is the ``Iverson bracket'' returning $1$ if the input statement is true ($0$, otherwise).

\STATE \textbf{Output} 
change detection map $\bC^\star$.
\end{algorithmic}
\end{algorithm}

The question is, however, how to estimate those unchanged locations $\ell$ specified by $\bm\Omega^{t+1}$.
Under the unsupervised setting, the best we could do is to measure the distance between $\bx_\ell$ and $\by_\ell$, to initiate the procedure.
As discussed in Section \ref{sec:problem}, Euclidean distance in \eqref{eq:DI} is not suitable given the fact that an unchanged location $\ell$ could have quite different $(\bx_\ell,\by_\ell)$ under diverse illumination conditions; we will show that Euclidean distance leads to poor HCD results (cf. Section \ref{sec:ablation}), motivating us to adopt the spectral angle mapper (SAM) as a more robust distance measure \cite{SAM}, i.e., 
\begin{equation}\label{eq: SAM}
\textrm{SAM}(\bx_\ell,\by_\ell)
\triangleq 
\arccos 
\left(
\frac{\bx_\ell^T\by_\ell}{\|\bx_\ell\|_2\|\by_\ell\|_2}
\right),
\end{equation}
which is, at least, not affected by multiplicative changes in image illumination conditions.
To see it, we have $\textrm{SAM}(\bx_\ell,\by_\ell)=\textrm{SAM}(\alpha\bx_\ell,\beta\by_\ell)$, where the meaning of $(\alpha,\beta)$ is any nonzero illumination condition (not hyperparameter).
Note that we do not need to estimate $(\alpha,\beta)$ from images, because their values do not affect the SAM value (hence are not needed in the HyperLUCID algorithm).
Furthermore, we safely argue that those pixels with the smallest SAM values are unchanged ones, as small SAM values indicate highly similar hyperspectral shapes between $(\bx_\ell,\by_\ell)$ [cf. \eqref{eq: SAM}] when ignoring the possible scaling of $(\alpha,\beta)$.
We will experimentally show that low-SAM pixels do correspond to unchanged regions with very high probability (over 99\% for many benchmark HCD datasets).
For safety, only a small portion of 20\% pixels with the smallest SAM values are collected into $\bm\Omega^1$ to initiate the procedure; this is reasonable, as most pixels in real benchmark HCD datasets $(\bX,\bY)$ are unchanged ones in typical real-world detection scenarios \cite{JocelynHAD,CDADMM}.

In order to meet condition (A), even though the other 80\% pixels may also contain unchanged ones, they are for safety not collected in $\bm\Omega^1$ in the beginning.
Instead, we first use $\bm\Omega^1$ to learn how to calibrate the spectra, thereby yielding $f^1$ (i.e., Line 6 in Algorithm \ref{algo:HyperLUCID}), and assume that by comparing the calibrated image $\bX^1:=f^1(\bX^0)$ with $\bY$, we can identify more unchanged pixels using the SAM distance measure in \eqref{eq: SAM}, thereby forming a union incremental set $\bm\Omega^2$ (i.e., Line 4 and Line 5 in Algorithm \ref{algo:HyperLUCID}).
By iteratively repeating the above procedure, from $\bm\Omega^t$ we gather more unchanged pixels to form $\bm\Omega^{t+1}$ for learning a more accurate hyperspectral calibration function, i.e., $f^{t+1}$, as detailed in Algorithm \ref{algo:HyperLUCID}.
This procedure is repeated until the difference of standard deviation of the SAM values (computed from $\bm\Omega^{t+1}$) is less than 0.05 compared with the last iteration $t$.\footnote{As a well-known mathematical result \cite{converge}, an upper-bounded sequence must converge.
Thus, as $\{\bm\Omega^t\}$ is increasing (cf. \eqref{eq:increasingOmega}) with a trivial finite upper bound, i.e., $|\bm\Omega^t|\leq L<\infty$, the sequence $\{\bm\Omega^t\}$ must converge, implying that the stopping criterion must also be satisfied within finite steps (cf. Figure \ref{fig:fivecurves}), thereby ensuring the convergence of HyperLUCID.}
Finally, the changed pixels can be detected by thresholding the SAM values computed from the calibrated image $\bX^t$ and $\bY$ (i.e., Line 10 in Algorithm \ref{algo:HyperLUCID}), thereby forming the final change detection map $\bC^\star\in\{0,1\}^L$ defined by Algorithm \ref{algo:HyperLUCID} with 1 (resp., 0) indicating changed pixels (resp., unchanged pixels).

We will experimentally demonstrate that condition (A) is almost satisfied under the proposed safe algorithmic scheme (cf. Section \ref{sec:discussion}).
In case that $\bm \Omega^{t+1}$ (and hence $\bm\Omega^\star$) contains some minor portion of changed pixels in $\bC$ (i.e., outliers), the adopted loss function \eqref{def:loss} elegantly introduces an outlier-robust mechanism as formalized in the following property.  
\begin{Property}\label{prop:L1robust}
The masked loss function \eqref{def:loss} is robust against the violation of condition (A), as those larger fitting errors brought by very few changed pixels (considered as outliers in $\bm \Omega^{t+1}$) will be elegantly ignored during the outlier-robust L1-norm optimization \cite{CVXbookCLL2016}.
\hfill$\square$
\end{Property}

\noindent
To concisely derive Property \ref{prop:L1robust}, define $\bv_\ell^{t+1}\triangleq f^{t+1}(\bx^t_\ell)-\by_\ell\in\mathbb{R}^M$.
So, \eqref{def:loss} can be reformulated as 
\begin{align}
\frac{1}{N^{t+1}} & \sum_{\ell\in\bm\Omega^{t+1}}\sum_{m=1}^M |[\bv_\ell^{t+1}]_m| 
=
\frac{1}{N^{t+1}}\sum_{m=1}^M \sum_{\ell\in\bm\Omega^{t+1}} |[\bv_\ell^{t+1}]_m| 
\nonumber
\\
&=
\frac{1}{N^{t+1}} \sum_{m=1}^M \left\| ([\bv_{\ell_1}^{t+1}]_m,\dots,[\bv_{\ell_{N^{t+1}} }^{t+1}]_m) \right\|_1, \label{def:lossnew}
\end{align} 
where $|\cdot|$ is the absolute value, $[\bv_\ell^{t+1}]_m$ denotes the $m$th entry of $\bv_\ell^{t+1}$, and $N^{t+1}\triangleq|\bm\Omega^{t+1}|$ denotes the number of elements in $\bm\Omega^{t+1}:=\{\ell_1,\dots,\ell_{N^{t+1}}\}$.
As a well-known property in the optimization area \cite{CVXbookCLL2016}, minimizing the L1-norm $\|\bv\|_1$ tends to return a sparse solution $\bv$.
This, together with the observations that $[\bv_{\ell n}^{t+1}]_m$ tends to be zero for unchanged pixels $\ell_n$ (i.e., the majority of $\bm\Omega^{t+1}$) and that $[\bv_{\ell_n}^{t+1}]_m$ tends to be large for changed pixels $\ell_n$ (i.e., outliers of $\bm\Omega^{t+1}$) as indicated by Property \ref{prop:GoodMapping}, implies that minimizing the L1-norm in \eqref{def:lossnew} (i.e., convex envelope of L0-norm \cite{CVXbookCLL2016}) yields a sparse solution, focusing on well fitting the majority of unchanged pixels while ignoring some large fitting errors brought by the minor portion of the changed pixels.

\begin{figure}[t]
\begin{center}
\includegraphics[width=0.49\textwidth]{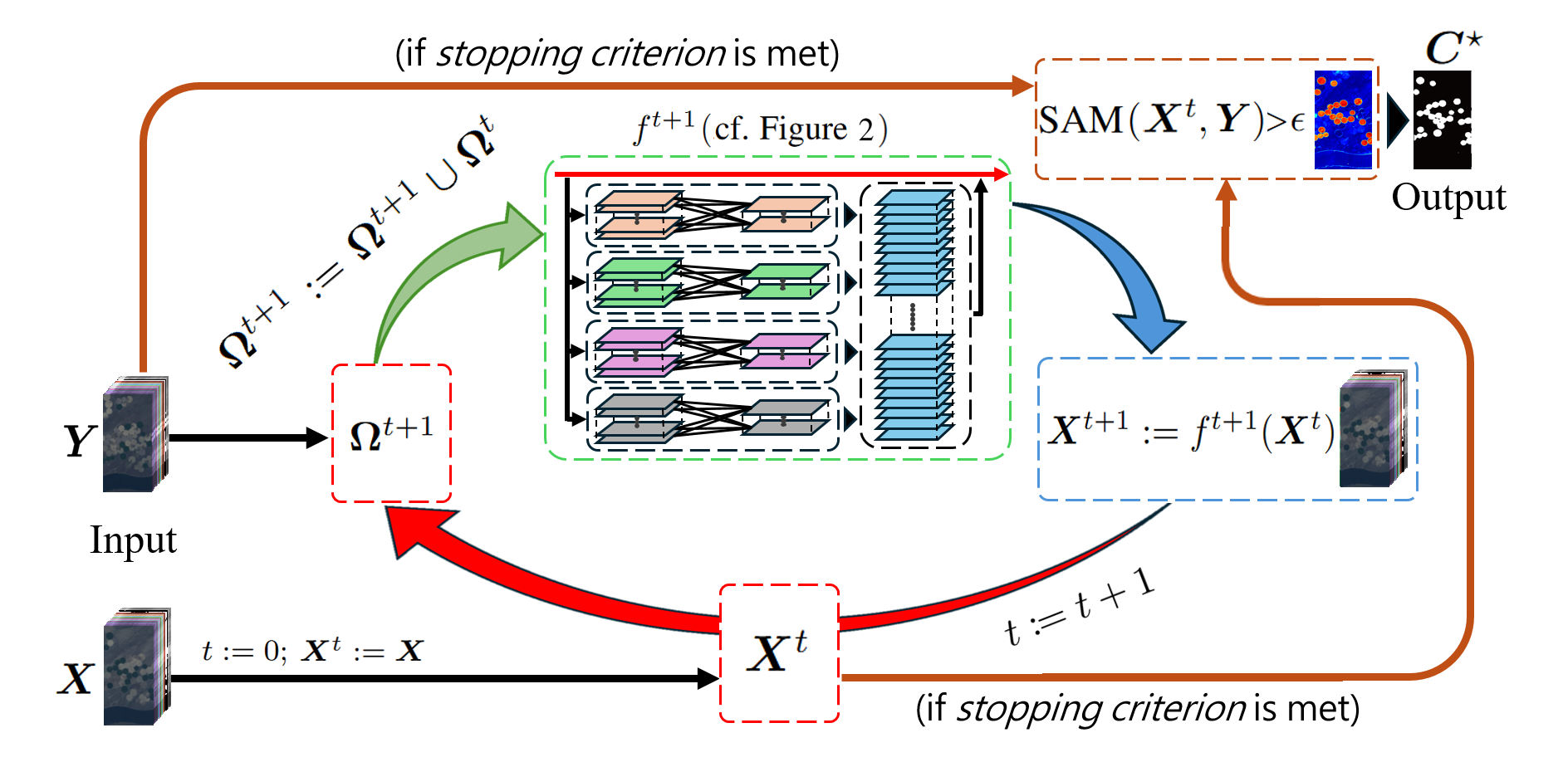}
\caption{As indicated by the red/green/blue arrows, the unsupervised HyperLUCID algorithm (i.e., Algorithm \ref{algo:HyperLUCID}) iteratively learns the calibration function $f^{t+1}$ using a gradually augmented training set $\bm\Omega^{t+1}$ of safely selected unchanged samples, in order to compensate the variability of acquisition conditions of $(\bX,\bY)$.
Once the predefined stopping criterion is met, the calibrated image $\bX^{t}$ (together with the input $\bY$) is sent to the HyperCAD-based change detector for computing the change map $\bC^\star$ [cf. \eqref{eq:DI-modify} and \eqref{eq: SAM}], as indicated by the brown arrows, thereby facilitating a simple yet effective detection procedure with state-of-the-art HCD performance.
}\label{fig:flow}
\end{center}
\end{figure}

We still need to discuss the aforementioned conditions (B) and (C).
One can see that condition (C) is clearly true from the iterative scheme proposed in Algorithm \ref{algo:HyperLUCID}, which iteratively estimates an increasing sequence of index sets of unchanged pixels (cf. Line 4 and Line 5 in Algorithm \ref{algo:HyperLUCID}).
As for condition (B), note that $\bU\setminus\bm\Omega^\star \approx \emptyset$ is equivalent to require that the set $\bm\Omega^\star:=\lim_{t\rightarrow\infty}\bm\Omega^t$ gets as large as possible [note that $\bm\Omega^\star$ is a subset of $\bU$ according to condition (A)].
Through the incremental set computing via the union mechanism (cf. Line 4 and Line 5 in Algorithm \ref{algo:HyperLUCID}), one can know that $\bm\Omega^t$ becomes larger as the algorithm evolves.
Thus, we know that $\bm\Omega^t$ will get closer and closer to $\bU$, and comprehensive real data experiments demonstrate that its limit $\bm\Omega^\star:=\lim_{t\rightarrow\infty}\bm\Omega^t$ does gradually approach $\bU$, as alluded by the state-of-the-art HCD performance of the proposed HyperLUCID algorithm, which is graphically illustrated in Figure \ref{fig:flow}.

We conclude this section by reporting some implementation details of HyperLUCID (i.e., Algorithm \ref{algo:HyperLUCID}) for reproducibility, together with some remarks.
The number of epochs at the $t$th iteration is empirically set as $100\times 0.7^t$ when training the network $f$ (cf. Figure \ref{fig:netf}).
This setting is coupled together with the warm start strategy, which well exploits the historical training experience of $f^t$ when learning $f^{t+1}$.
Specifically, the network parameters in $f^t$ will be used to initialize the network parameters of $f^{t+1}$ for fast convergence at the $(t+1)$th iteration, thereby allowing us to decrease the number of epochs as stated above.
Also, when learning $f^{t+1}$, we have ensured that the training data in $\bm\Omega^{t+1}$ are fully used in order to yield an iteratively upgraded hyperspectral calibration function $f^{t+1}$.
The network architecture $f$ (cf. Figure \ref{fig:netf}) uniformly divides the input hyperspectral bands into four groups in the first phase, and in the second phase the number of CNN channels is always unified as 64 across different datasets regardless of $M$.
One remarkable property of HyperLUCID is that all its parameters/hyperparameters reported in this section are unified (data-independent) for uniformly yielding superior HCD results on several benchmark datasets, implying a highly user-friendly HCD tool.
Another remarkable property is that the unsupervised HyperLUCID algorithm even outperforms most existing semi-supervised HCD methods.

\vspace{-0.1cm}
\section{Experimental Results}\label{sec: experiment}
In this section, we demonstrate the effectiveness of the proposed HyperLUCID on real hyperspectral datasets by comparing it against the state-of-the-art HCD algorithms. 
Section \ref{sec:dataset} introduces the five real-world hyperspectral datasets used in our experiments.
The benchmark HCD methods and the definitions of the five quantitative metrics are detailed in Section \ref{sec:exp_set}, together with the experimental setup.
In Section \ref{sec:exp_analysis}, we present a comprehensive performance analysis of HyperLUCID compared to the benchmark HCD algorithms.
Section \ref{sec:ablation} presents an ablation study to validate the contribution of each customized module (i.e., residual, SAM and CNN model) in HyperLUCID.
Finally, comprehensive discussions are collectively presented in Section \ref{sec:discussion}.

\begin{figure}[t]
\centering
\subfigure[]{
\includegraphics[width=0.145\textwidth]{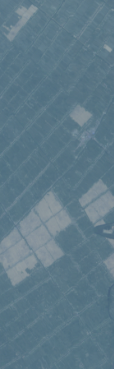}
}
\subfigure[]{
\includegraphics[width=0.145\textwidth]{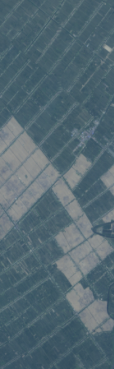}
}
\subfigure[]{
\includegraphics[width=0.145\textwidth]{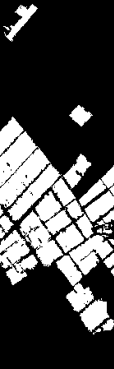}
}
\caption{False-color images $(\bX,\bY)$ and the GT change map of the Yancheng dataset. (a) HSI image captured on May 3rd, 2006. (b) HSI image captured on April 23rd, 2007. (c) GT change map, where changed pixels are shown in white and unchanged pixels in black.
}
\label{fig:pseudo-color_Farm}

\end{figure}

\begin{figure}[t]
\centering
\subfigure[]{
\includegraphics[width=0.145\textwidth]{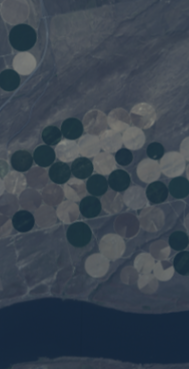}
}
\subfigure[]{
\includegraphics[width=0.145\textwidth]{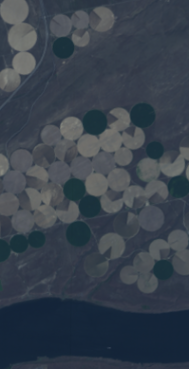}
}
\subfigure[]{
\includegraphics[width=0.145\textwidth]{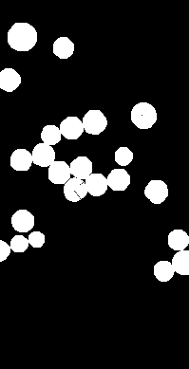}
}
\caption{False-color images $(\bX,\bY)$ and the GT change map of the Hermiston dataset. (a) HSI image captured on May 1st, 2004. (b) HSI image captured on May 8th, 2007. (c) GT change map, where changed pixels are shown in white and unchanged pixels in black.
}
\label{fig:pseudo-color_Hermiston}
\end{figure}

\begin{figure}[t]
\centering
\subfigure[]{
\includegraphics[width=0.145\textwidth]{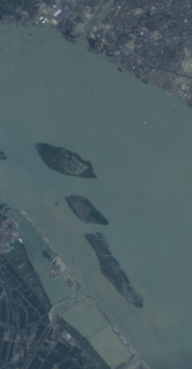}
}
\subfigure[]{
\includegraphics[width=0.145\textwidth]{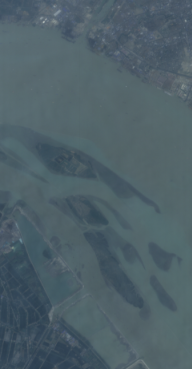}
}
\subfigure[]{
\includegraphics[width=0.145\textwidth]{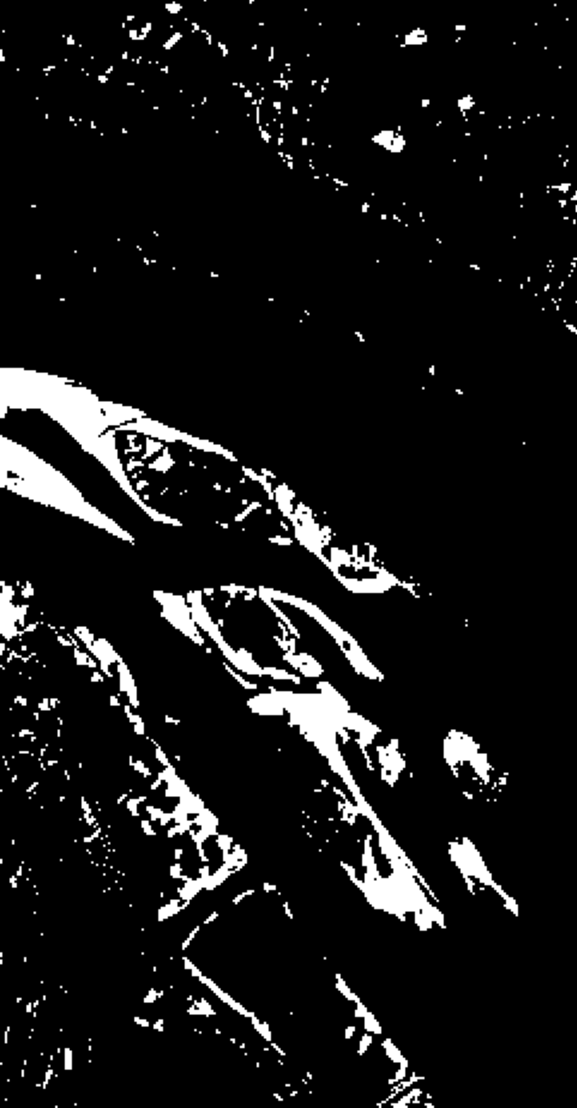}
}
\caption{False-color images $(\bX,\bY)$ and the GT change map of the Jiangsu dataset. (a) HSI image captured on May 3rd, 2013. (b) HSI image captured on December 31st, 2007. (c) GT change map, where changed pixels are shown in white and unchanged pixels in black.
}
\label{fig:pseudo-color_river}
\end{figure}

\begin{figure}[t]
\centering
\subfigure[]{
\includegraphics[width=0.145\textwidth]{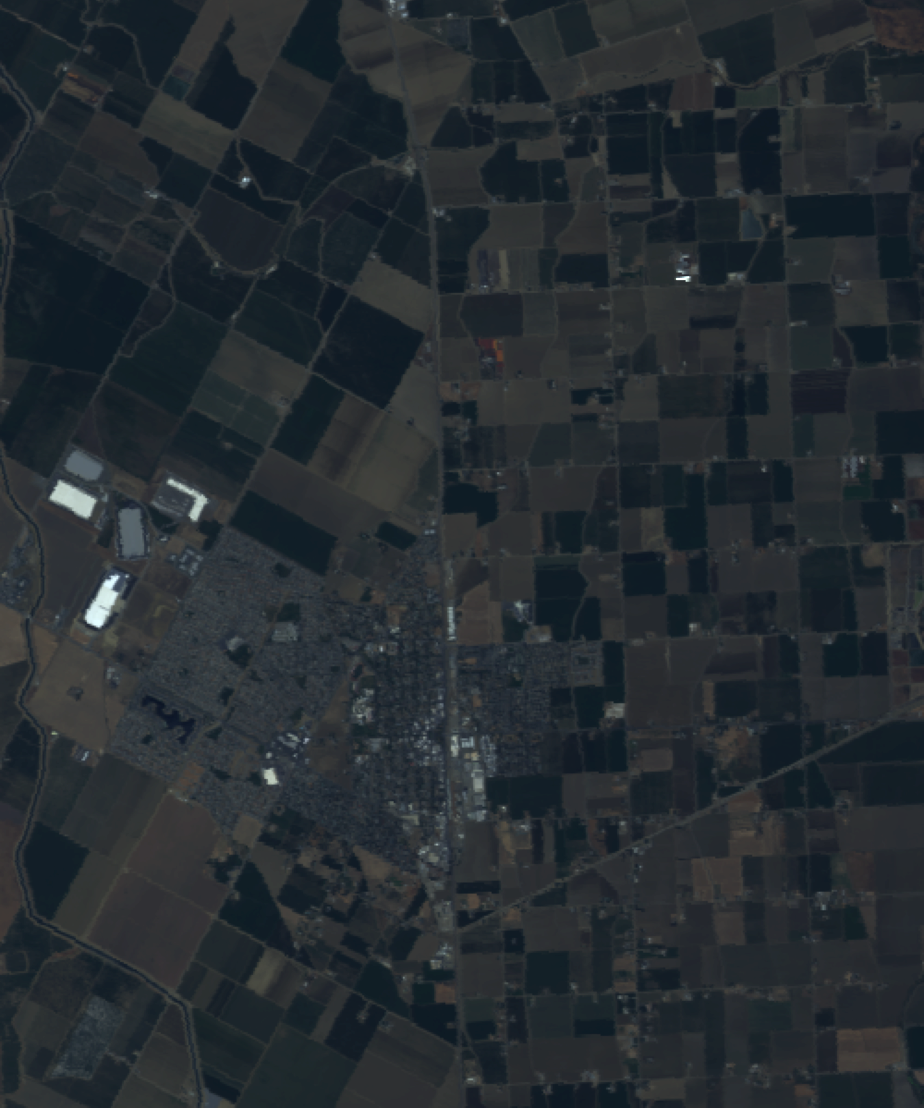}
}
\subfigure[]{
\includegraphics[width=0.145\textwidth]{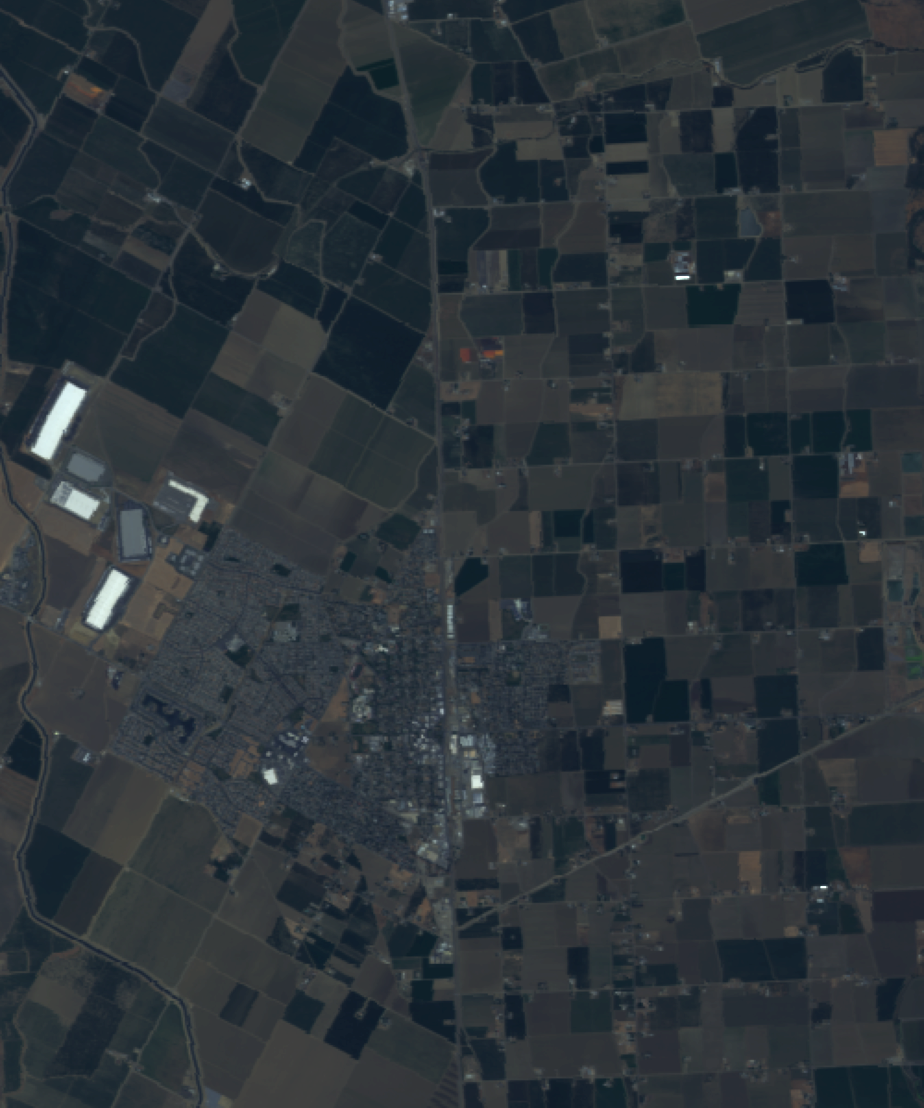}
}
\subfigure[]{
\includegraphics[width=0.145\textwidth]{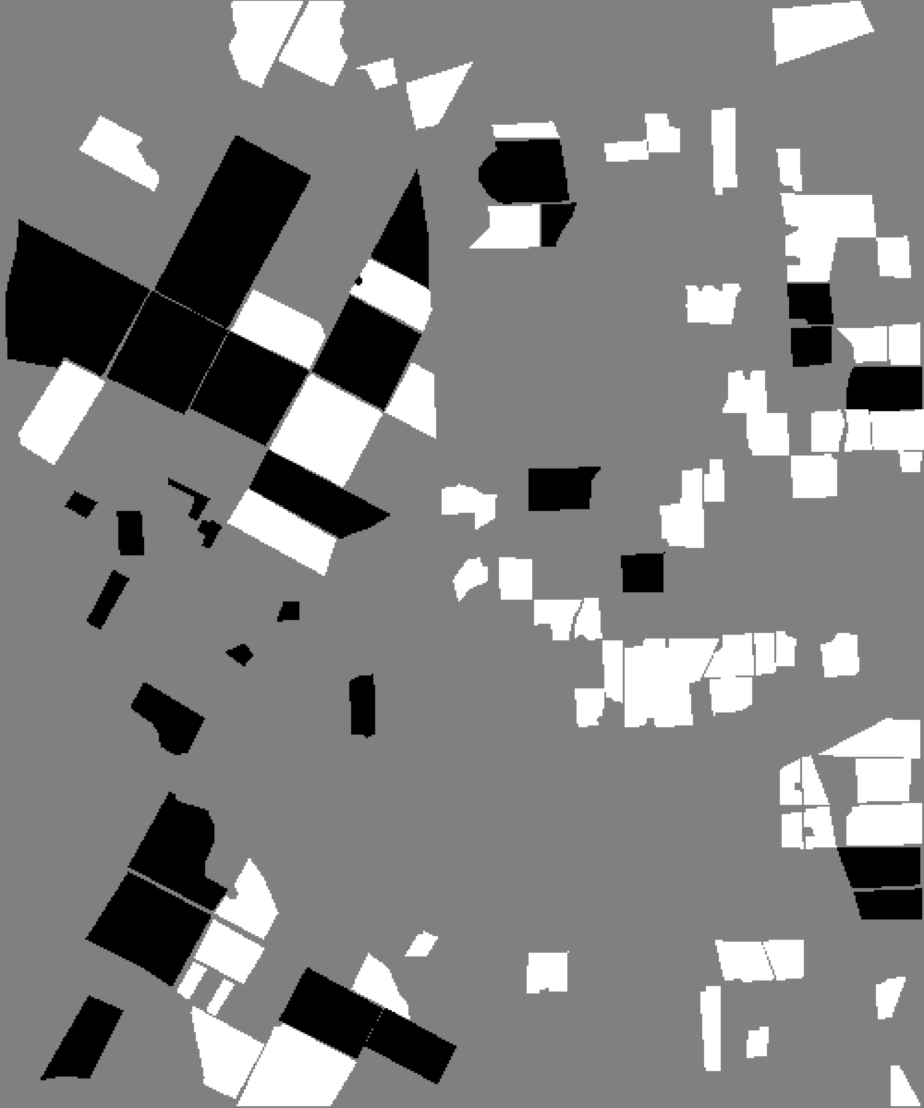}
}
\caption{False-color images $(\bX,\bY)$ and the GT change map of the Bay Area dataset. (a) HSI image captured in 2013. (b) HSI image captured in 2015. (c) GT change map, where changed, unchanged, and unknown pixels are represented in white, black, and grey, respectively.
}
\label{fig:pseudo-color_Bay}
\end{figure}

\begin{figure}[t]
\centering
\subfigure[]{
\includegraphics[width=0.145\textwidth]{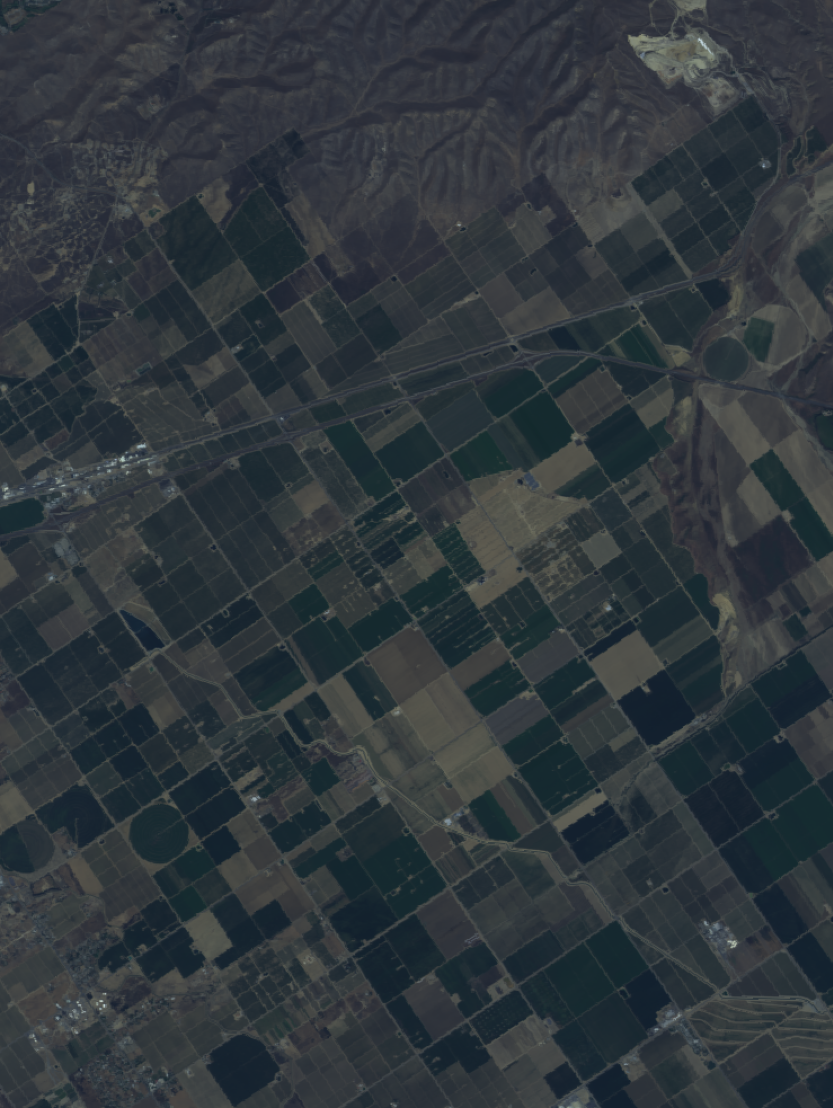}
}
\subfigure[]{
\includegraphics[width=0.145\textwidth]{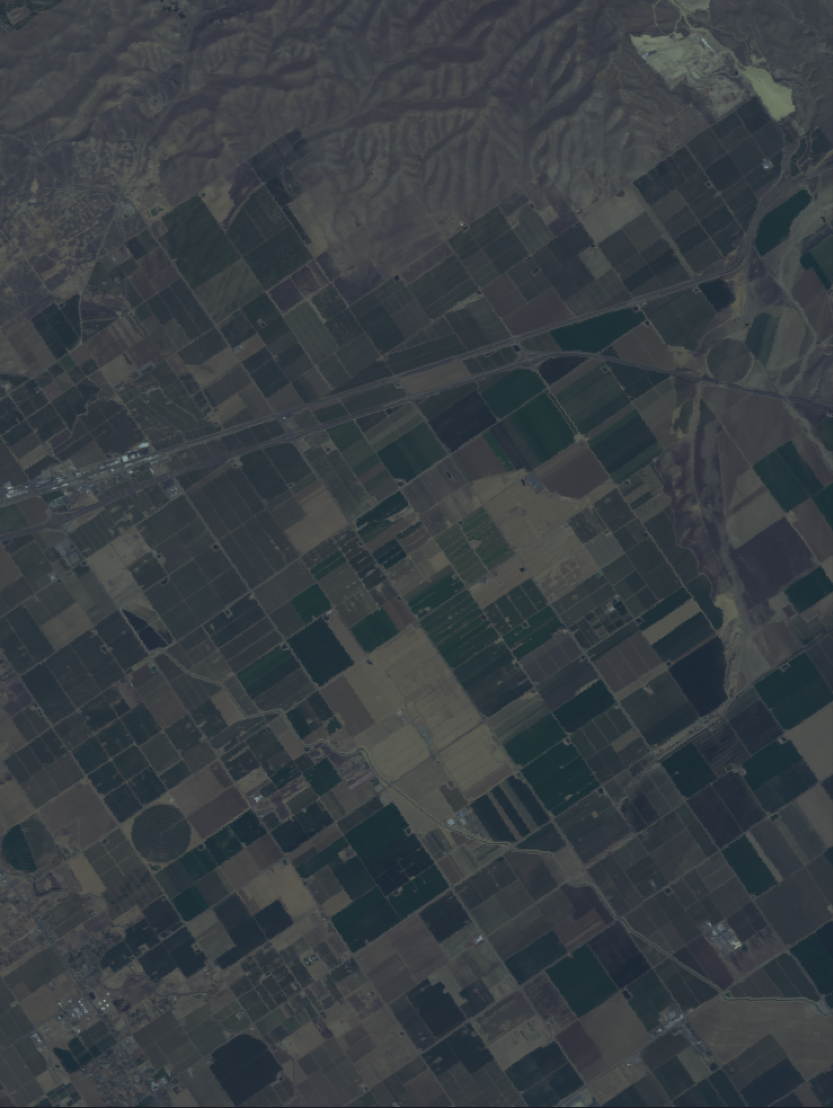}
}
\subfigure[]{
\includegraphics[width=0.145\textwidth]{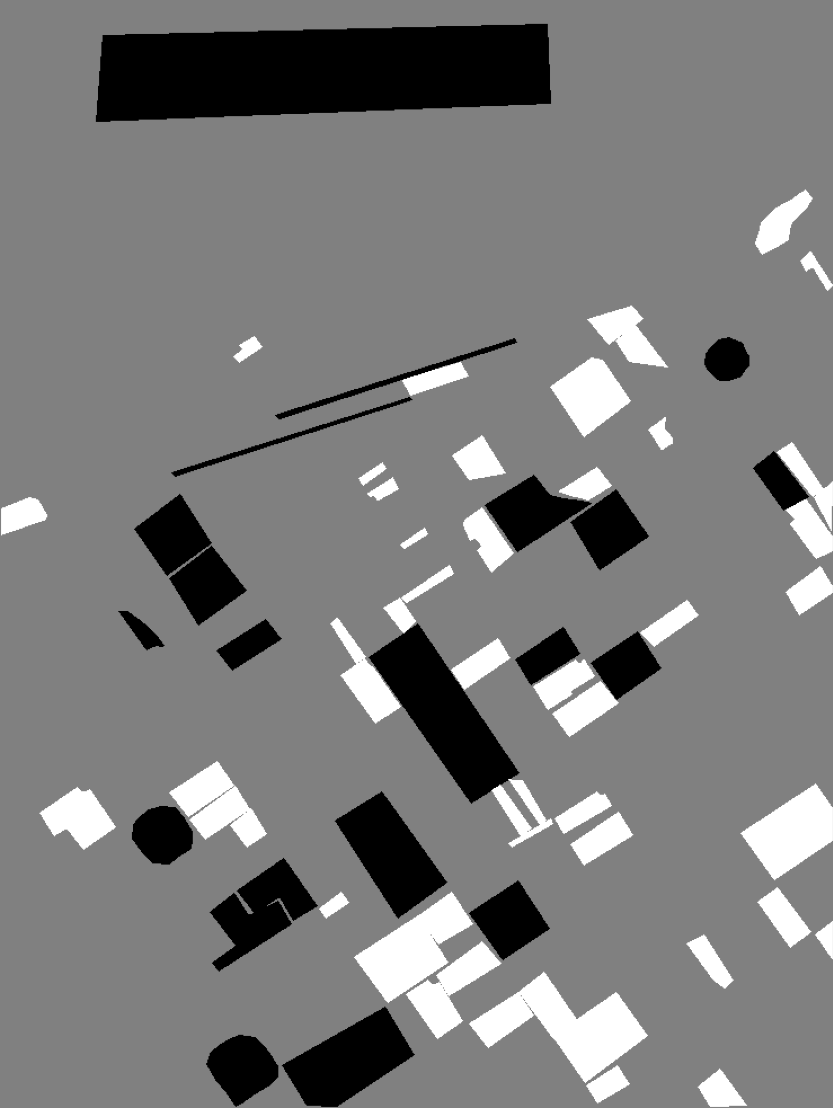}
}
\caption{False-color images $(\bX,\bY)$ and the GT change map of the Santa Barbara dataset. (a) HSI image captured in 2013. (b) HSI image captured in 2014. (c) GT change map, where changed, unchanged, and unknown pixels are represented in white, black, and grey, respectively.
}
\label{fig:pseudo-color_Barbara}
\end{figure}

\subsection{Dataset Descriptions}\label{sec:dataset}

The Yancheng dataset was captured by the Hyperion sensor over a wetland agricultural area in Yancheng, Jiangsu Province, China \cite{Yangcheng}. 
After noise removal, the bitemporal HSI images contain $M=155$ spectral bands with $450\times 140$ pixels \cite{QUEENG}.
This dataset mainly captures agricultural changes, such as shifts in vegetation, bare land, water, and soil \cite{liu2019unsupervised}.
The false-color images and the corresponding GT change map are shown in Figure \ref{fig:pseudo-color_Farm}.

The Hermiston dataset was captured by the Hyperion sensor over an agricultural area in Hermiston, Umatilla County, United States, with $M=242$ spectral bands and $390\times 200$ pixels \cite{guo2021change}.
The change areas in this dataset are mainly caused by variations in crops, water, and soil.
The false-color images and the corresponding GT change map are shown in Figure \ref{fig:pseudo-color_Hermiston}.

The Jiangsu dataset was captured by the Hyperion sensor over a river area in Jiangsu Province, China \cite{wang2018getnet}.
After noise removal, the bitemporal HSI images contain $M= 198$ spectral bands with $463 \times 241$ pixels \cite{guo2021change}.
Most of the detected changes are due to variations in materials associated with river course changes over time.
The false-color images and the corresponding GT change map are shown in Figure \ref{fig:pseudo-color_river}.

The Bay Area dataset was captured by the Airborne Visible/Infrared Imaging Spectrometer (AVIRIS) sensor over an agricultural area in Patterson, California, United States, with $M=224$ spectral bands and $600\times 500$ pixels \cite{HyperNet}.
This dataset primarily captures changes associated with farmland and buildings.
The false-color images and the corresponding GT change map are shown in Figure \ref{fig:pseudo-color_Bay}.

The Santa Barbara dataset was captured by the AVIRIS sensor over an agricultural area in Santa Barbara, California, United States, with $M=224$ spectral bands and $984 \times 740$ pixels \cite{HyperNet}.
The dataset mainly reflects urban development and dynamic changes in farmland.
The false-color images and the corresponding GT change map are shown in Figure \ref{fig:pseudo-color_Barbara}.

\subsection{Experimental Settings}\label{sec:exp_set}
To comprehensively assess the effectiveness of the proposed HyperLUCID, eight state-of-the-art change detection methods are selected for baseline comparisons, including HyperNet \cite{HyperNet}, $\text{S}^3$Net \cite{S3Net}, QUEEN-$\mathcal{G}$ \cite{QUEENG}, adaptive center-focused hybrid attention network (ACFHAN) \cite{ACFHAN}, 
CVA \cite{bovolo2006theoretical}, PCA-Kmeans \cite{PCAKMeans}, TDRD \cite{TDRD}, and PTCD \cite{PTCD}.
% Quantitative metrics
To evaluate the performance of HCD, we adopt five quantitative metrics: overall accuracy (OA) \cite{nishii2002accuracy}, Kappa coefficient ($\kappa$) \cite{brennan1981coefficient}, F1-score (F1) \cite{sathyanarayanan2024confusion}, precision (Pr) \cite{sathyanarayanan2024confusion}, and recall (Re) \cite{sathyanarayanan2024confusion}.
These quantitative metrics are derived from confusion matrix elements, including true positive (TP), true negative (TN), false positive (FP), and false negative (FN).
Each metric is separately introduced in detail below.

OA represents the ratio of correctly classified pixels to the total number of pixels, providing an intuitive measure of classification accuracy, which is defined as
$
\text{OA} = \frac{\text{TP} + \text{TN}}{\text{TP} + \text{TN} + \text{FP} + \text{FN}}.
$
The measure $\kappa$ offers a more reliable assessment than OA by accounting for inter-rater reliability \cite{brennan1981coefficient}, which is defined as
\begin{align*}
\kappa = \frac{\text{OA} - \text{Pe}}{1 - \text{Pe}},
\end{align*}
where $\text{Pe}$ is the hypothetical probability of random agreement, defined as
$
\text{Pe} = \frac{(\text{TP} + \text{FP})(\text{TP} + \text{FN}) + (\text{FP} + \text{TN})(\text{FN} + \text{TN})}{(\text{TP} + \text{TN} + \text{FP} + \text{FN})^2}.
$
Similarly, Pr, Re, and F1 are adopted as complementary metrics to OA, especially for handling class imbalance scenarios.
Pr and Re are defined as $\text{Pr} = \frac{\text{TP}}{\text{TP} + \text{FP}}$ and $\text{Re} = \frac{\text{TP}}{\text{TP} + \text{FN}}$, respectively, and their joint measure is known as F1, defined as
\begin{align*}
\text{F1} = 2\times\frac{\text{Pr}\times\text{Re}}{\text{Pr} + \text{Re}},
\end{align*}
serving as a comprehensive evaluation combining Pr and Re.

For the hyperspectral calibration network in the proposed HyperLUCID, the model training is optimized using the Adam optimizer \cite{kingma2014adam} with an initial learning rate of 0.005, which decays by a factor of 0.9 every 20 epochs.
L1 loss \cite{janocha2017loss} is used as the loss function for the proposed HyperLUCID.
Training is performed iteratively with warm starts between successive runs, meaning that each run initializes from the previous model's parameters to ensure continuity and accelerate convergence. 
The first iteration is trained for 100 epochs, and in each subsequent iteration, the number of training epochs is reduced by a factor of 0.7 until the predefined stopping criterion is met.
To reduce variance from the stochastic nature of deep learning, the final results are reported as average performance over 10 independent Monte Carlo trials. 

To adapt the HSIs dataset for $\text{S}^3$Net, we first selected the red, green, and blue bands according to the wavelength range of each dataset.
Then, we applied summation and average operations to compress the respective red, green, and blue bands.
The region sizes of simple linear iterative clustering (SLIC) \cite{SLIC} are set to 13, 10, 14, 16, and 14, corresponding to the Yancheng, Hermiston, Jiangsu, Bay, and Barbara datasets, respectively, to ensure optimal performance for each dataset.
Additionally, we set the segmentation scale parameter to 6 for all datasets, which was utilized to extract patch pairs for model training.
On the other hand, since the original TDRD and PTCD papers do not specify a threshold selection strategy for generating change maps, we adopt a statistical approach by setting the threshold as the sum of the mean and standard deviation of the detection output \cite{sezgin2004survey}. 
This approach provides a reasonable balance between Pr and Re.
All experiments, except for the edge-computing experiment in Section \ref{sec:discussion}, are conducted using Python 3.11.10 and PyTorch 2.5.1 on a computer equipped with an Intel Core i9-10900K CPU and an NVIDIA GeForce RTX 3090 GPU.
The reported running time for each method corresponds to its algorithmic execution time measured on the above computing environment.

\vspace{-0.1cm}
\subsection{Quantitative and Qualitative Analysis}\label{sec:exp_analysis}

\begin{table*}[t]
\centering
\scriptsize
\caption{Quantitative comparisons on five real benchmark HCD datasets. The results of the quantum algorithm, QUEEN-$\mathcal{G}$ (semi-supervised), on the first three datasets are from \cite[Table II]{QUEENG}. The boldfaced underlined number indicates the best performance, while the boldfaced number indicates the second-best. HyperLUCID shows the best performance among the unsupervised methods, and demonstrates high computational efficiency. 
Results across 10 Monte Carlo runs are reported as mean $\pm$ standard deviation (std); zero std is omitted to save space.
The computational time is reported in seconds.
}
\label{tab:results}
\setlength{\tabcolsep}{1.55pt}
\renewcommand{\arraystretch}{1.18}

\newcommand{\best}[1]{$\underline{\smash{\mathbf{#1}}}$}
\newcommand{\second}[1]{$\mathbf{#1}$}
\begin{tabular}{c|c|cccc|ccccc}
\hline
\multicolumn{2}{c|}{} 
& \multicolumn{4}{c|}{Other Methods}
& \multicolumn{5}{c}{Unsupervised Methods} \\
\hline
Dataset & Index & HyperNet & S$^3$Net & QUEEN-$\mathcal{G}$ & ACFHAN & CVA & PCA-Kmeans & TDRD & PTCD & HyperLUCID \\
\hline

\multirow{6}{*}{Yancheng}
& OA$\uparrow$ & $0.585\pm0.183$ & $0.931\pm0.001$ & \best{0.969}$\pm0.006$ & $0.933\pm0.003$ & $0.960$ & \second{0.961} & $0.945$ & $0.923$ & \best{0.969}$\pm0.001$ \\
& $\kappa\uparrow$ & $0.314\pm0.248$ & $0.842\pm0.002$ & \second{0.925}$\pm0.014$ & $0.827\pm0.009$ & $0.905$ & $0.906$ & $0.860$ & $0.803$ & \best{0.926}$\pm0.003$ \\
& F1$\uparrow$ & $0.598\pm0.128$ & $0.892\pm0.002$ & \best{0.947}$\pm0.010$ & $0.872\pm0.008$ & \second{0.933} & \second{0.933} & $0.897$ & $0.854$ & \best{0.947}$\pm0.002$ \\
& Pr$\uparrow$ & $0.461\pm0.165$ & $0.819\pm0.003$ & $0.926$$\pm0.028$ & \second{0.974}$\pm0.003$ & $0.920$ & $0.929$ & \best{0.985} & $0.950$ & $0.948\pm0.002$ \\
& Re$\uparrow$ & $0.944\pm0.032$ & \best{0.984}$\pm0.002$ & \second{0.970}$\pm0.013$ & $0.789\pm0.014$ & $0.946$ & $0.937$ & $0.823$ & $0.776$ & $0.947\pm0.005$ \\
& Time$\downarrow$ & $18.125\pm0.598$ & $31.455\pm7.231$ & $206.090\pm4.026$ & $11688.382\pm7.480$ & \best{0.044} & $30.086$ & $4.545$ & $87.552$ & \second{3.576}$\pm0.162$ \\
\hline

\multirow{6}{*}{Hermiston}
& OA$\uparrow$ & $0.937\pm0.018$ & $0.938\pm0.003$ & \best{0.986}$\pm0.004$ & $0.973\pm0.001$ & $0.977$ & $0.965$ & $0.958$ & \second{0.979} & \second{0.979}$\pm0.000$ \\
& $\kappa\uparrow$ & $0.767\pm0.057$ & $0.758\pm0.012$ & \best{0.937}$\pm0.017$ & $0.871\pm0.006$ & $0.899$ & $0.852$ & $0.830$ & $0.907$ & \second{0.912}$\pm0.000$ \\
& F1$\uparrow$ & $0.803\pm0.047$ & $0.793\pm0.010$ & \best{0.945}$\pm0.014$ & $0.885\pm0.006$ & $0.912$ & $0.872$ & $0.854$ & $0.919$ & \second{0.923}$\pm0.000$ \\
& Pr$\uparrow$ & $0.680\pm0.068$ & $0.695\pm0.017$ & \second{0.905}$\pm0.029$ & \best{0.985}$\pm0.002$ & $0.890$ & $0.811$ & $0.772$ & $0.902$ & $0.881\pm0.000$ \\
& Re$\uparrow$ & \second{0.984}$\pm0.003$ & $0.923\pm0.001$ & \best{0.990}$\pm0.004$ & $0.804\pm0.010$ & $0.936$ & $0.943$ & $0.956$ & $0.937$ & $0.970\pm1.648$ \\
& Time$\downarrow$ & $20.575\pm0.502$ & $57.359\pm16.982$ & $253.640\pm1.648$ & $14521.071\pm13.996$ & \best{0.054} & $34.051$ & $8.643$ & $91.817$ & \second{1.626}$\pm0.003$ \\
\hline

\multirow{6}{*}{Jiangsu}
& OA$\uparrow$ & $0.922\pm0.024$ & $0.900\pm0.024$ & $0.943$$\pm0.011$ & $0.946\pm0.003$ & $0.912$ & \second{0.949} & $0.936$ & $0.875$ & \best{0.960}$\pm0.000$ \\
& $\kappa\uparrow$ & $0.579\pm0.076$ & $0.502\pm0.029$ & \second{0.706}$\pm0.037$ & $0.535\pm0.029$ & $0.610$ & $0.691$ & $0.670$ & $0.415$ & \best{0.713}$\pm0.002$ \\
& F1$\uparrow$ & $0.621\pm0.064$ & $0.555\pm0.023$ & \best{0.736}$\pm0.032$ & $0.559\pm0.027$ & $0.655$ & $0.719$ & $0.705$ & $0.480$ & \second{0.734}$\pm0.002$ \\
& Pr$\uparrow$ & $0.566\pm0.113$ & $0.453\pm0.022$ & $0.608$$\pm0.056$ & \best{0.962}$\pm0.014$ & $0.496$ & $0.697$ & $0.585$ & $0.376$ & \second{0.871}$\pm0.006$ \\
& Re$\uparrow$ & $0.708\pm0.030$ & $0.717\pm0.027$ & \second{0.941}$\pm0.034$ & $0.393\pm0.036$ & \best{0.964} & $0.742$ & $0.887$ & $0.665$ & $0.633\pm0.003$ \\
& Time$\downarrow$ & $25.291\pm0.476$ & $127.348\pm28.742$ & $360.290\pm8.703$ & $20690.346\pm98.742$ & \best{0.063} & $43.755$ & $10.740$ & $229.696$ & \second{4.162}$\pm0.059$ \\
\hline

\multirow{6}{*}{Bay Area}
& OA$\uparrow$ & $0.922\pm0.008$ & $0.750\pm0.001$ & \best{0.945}$\pm0.010$ & $0.892\pm0.003$ & $0.829$ & $0.786$ & $0.723$ & $0.675$ & \second{0.943}$\pm0.000$ \\
& $\kappa\uparrow$ & $0.843\pm0.016$ & $0.506\pm0.001$ & \second{0.891}$\pm0.018$ & $0.785\pm0.010$ & $0.663$ & $0.577$ & $0.463$ & $0.373$ & \best{0.885}$\pm0.001$ \\
& F1$\uparrow$ & $0.928\pm0.007$ & $0.732\pm0.002$ & \best{0.947}$\pm0.009$ & $0.893\pm0.007$ & $0.819$ & $0.774$ & $0.656$ & $0.575$ & \second{0.946}$\pm0.000$ \\
& Pr$\uparrow$ & $0.917\pm0.013$ & $0.857\pm0.001$ & \best{0.999}$\pm0.001$ & $0.950\pm0.009$ & $0.945$ & $0.888$ & \second{0.975} & $0.955$ & $0.953\pm0.001$ \\
& Re$\uparrow$ & \best{0.939}$\pm0.010$ & $0.639\pm0.001$ & \second{0.900}$\pm0.018$ & $0.842\pm0.013$ & $0.723$ & $0.686$ & $0.494$ & $0.411$ & \best{0.939}$\pm0.001$ \\
& Time$\downarrow$ & $94.355\pm0.712$ & $325.586\pm34.918$ & $950.373\pm2.321$ & $58965.395\pm42.108$& \best{0.358} & $283.308$ & $39.102$ & $3901.657$ &\second{19.227}$\pm1.855$\\
\hline

\multirow{6}{*}{Santa Barbara}
& OA$\uparrow$ & \second{0.915}$\pm0.009$ & $0.776\pm0.015$ & $0.837\pm0.027$ & $0.909\pm0.010$ & $0.841$ & $0.732$ & $0.752$ & $0.777$ & \best{0.936}$\pm0.002$ \\
& $\kappa\uparrow$ & \second{0.822}$\pm0.019$ & $0.517\pm0.028$ & $0.639\pm0.066$ & $0.807\pm0.021$ & $0.651$ & $0.496$ & $0.464$ & $0.496$ & \best{0.865}$\pm0.005$ \\
& F1$\uparrow$ & \second{0.893}$\pm0.011$ & $0.689\pm0.016$ & $0.748\pm0.051$ & $0.879\pm0.017$ & $0.769$ & $0.718$ & $0.656$ & $0.642$ & \best{0.916}$\pm0.004$ \\
& Pr$\uparrow$ & $0.878\pm0.017$ & $0.759\pm0.012$ & \best{0.998}$\pm0.001$ & $0.927\pm0.006$ & $0.894$ & $0.874$ & $0.721$ & $0.874$ & \second{0.955}$\pm0.003$ \\
& Re$\uparrow$ & \best{0.909}$\pm0.006$ & $0.631\pm0.029$ & $0.598\pm0.067$ & $0.835\pm0.011$ & $0.675$ & $0.610$ & $0.602$ & $0.507$ & \second{0.880}$\pm0.008$ \\
& Time$\downarrow$ & $230.767\pm0.811$ & $359.050\pm8.742$ & $6467.533\pm1.093$ & $146380.670\pm38.162$ & \best{0.711} & $633.453$ & $126.076$ & $14068.104$ & \second{50.626}$\pm13.172$\\
\hline

\end{tabular}
\end{table*}

\begin{figure*}[!t]
\centering
\subfigure[]{
\includegraphics[width=0.08\textwidth]{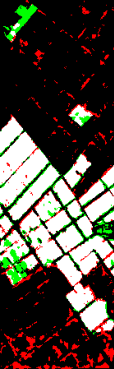}
}\hspace{-0.25cm}
\subfigure[]{
\includegraphics[width=0.081\textwidth]{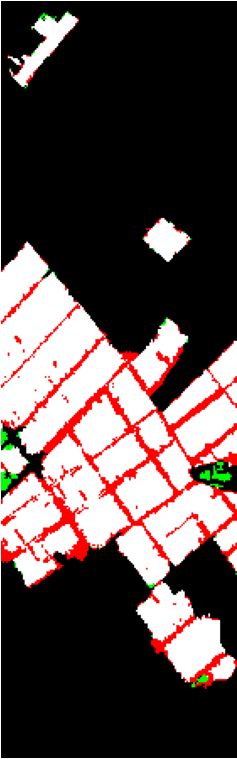}
}\hspace{-0.25cm}
\subfigure[]{
\includegraphics[width=0.08\textwidth]{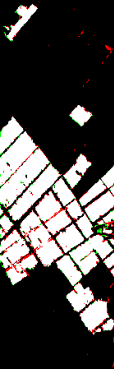}
}\hspace{-0.25cm}
\subfigure[]{
\includegraphics[width=0.0805\textwidth]{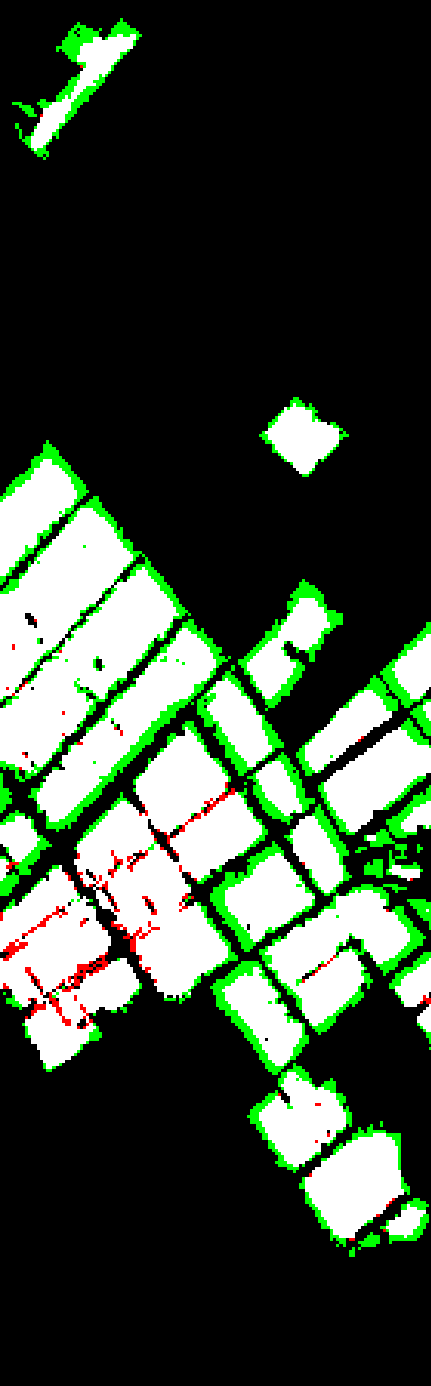}
}\hspace{-0.25cm}
\subfigure[]{
\includegraphics[width=0.08\textwidth]{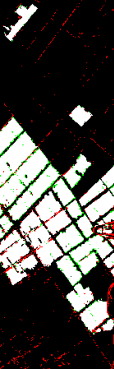}
}\hspace{-0.25cm}
\subfigure[]{
\includegraphics[width=0.08\textwidth]{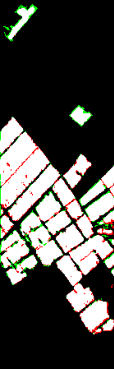}
}\hspace{-0.25cm}
\subfigure[]{
\includegraphics[width=0.08\textwidth]{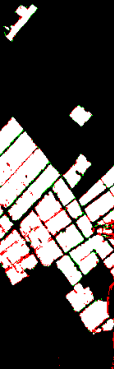}
}\hspace{-0.25cm}
\subfigure[]{
\includegraphics[width=0.08\textwidth]{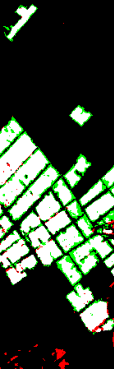}
}\hspace{-0.25cm}
\subfigure[]{
\includegraphics[width=0.08\textwidth]{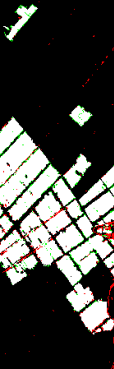}
}\hspace{-0.25cm}
\subfigure[]{
\includegraphics[width=0.0801\textwidth]{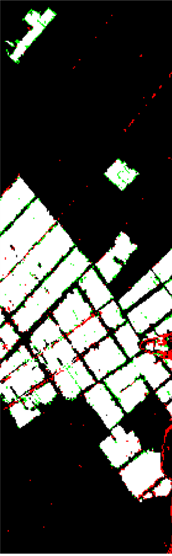}
}\hspace{-0.25cm}
\subfigure[]{
\includegraphics[width=0.08\textwidth]{gt_Farm}
}
\caption{Change detection results on the Yancheng dataset, where FP and FN are marked in red and green, respectively. (a) HyperNet. (b) $\text{S}^3$Net. (c) QUEEN-$\mathcal{G}$. (d) ACFHAN. (e) CVA. (f) PCA-Kmeans. (g) TDRD. (h) PTCD. (i) HyperLUCID. (j) HyperLUCID (NVIDIA Orin Nano). (k) GT.
}
\label{fig:results_Farm}
\end{figure*}

\begin{figure*}[!t]
\centering
\subfigure[]{
\includegraphics[width=0.08\textwidth]{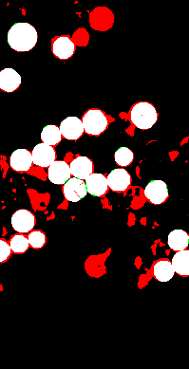}
}\hspace{-0.25cm}
\subfigure[]{
\includegraphics[width=0.0805\textwidth]{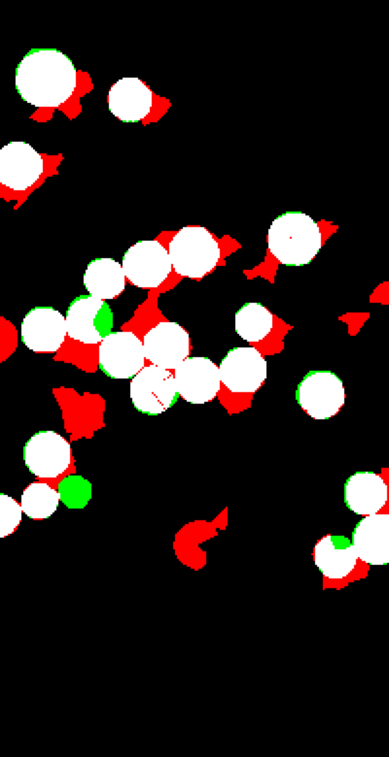}
}\hspace{-0.25cm}
\subfigure[]{
\includegraphics[width=0.08\textwidth]{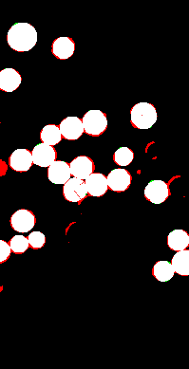}
}\hspace{-0.25cm}
\subfigure[]{
\includegraphics[width=0.0801\textwidth]{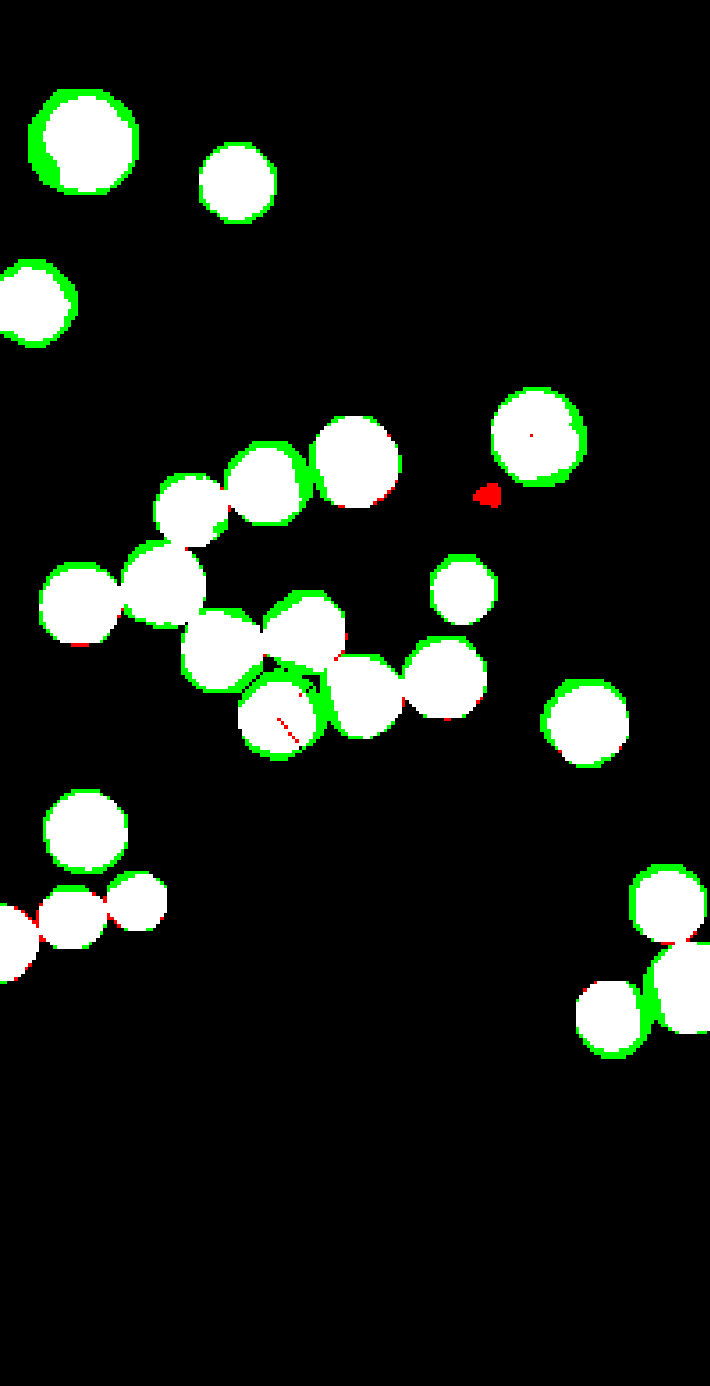}
}\hspace{-0.25cm}
\subfigure[]{
\includegraphics[width=0.08\textwidth]{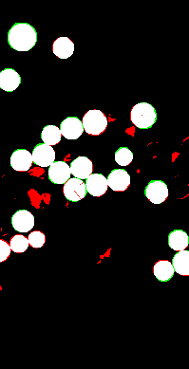}
}\hspace{-0.25cm}
\subfigure[]{
\includegraphics[width=0.08\textwidth]{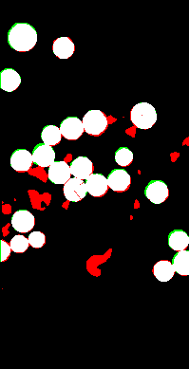}
}\hspace{-0.25cm}
\subfigure[]{
\includegraphics[width=0.08\textwidth]{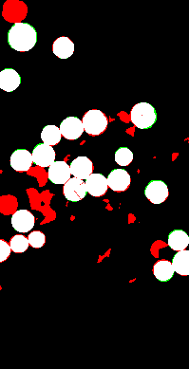}
}\hspace{-0.25cm}
\subfigure[]{
\includegraphics[width=0.08\textwidth]{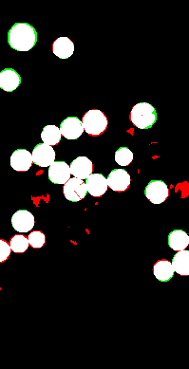}
}\hspace{-0.25cm}
\subfigure[]{
\includegraphics[width=0.08\textwidth]{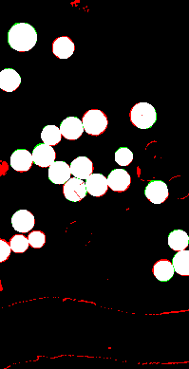}
}\hspace{-0.25cm}
\subfigure[]{
\includegraphics[width=0.081\textwidth]{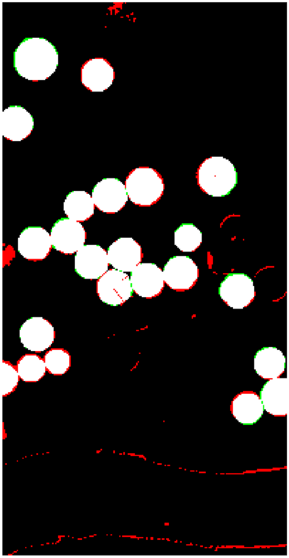}
}\hspace{-0.25cm}
\subfigure[]{
\includegraphics[width=0.08\textwidth]{gt_Hermiston}
}
\caption{Change detection results on the Hermiston dataset, where FP and FN are marked in red and green, respectively. (a) HyperNet. (b) $\text{S}^3$Net. (c) QUEEN-$\mathcal{G}$. (d) ACFHAN. (e) CVA. (f) PCA-Kmeans. (g) TDRD. (h) PTCD. (i) HyperLUCID. (j) HyperLUCID (NVIDIA Orin Nano). (k) GT.
}
\label{fig:results_Hermiston}
\end{figure*}

\begin{figure*}[!t]
\centering
\subfigure[]{
\includegraphics[width=0.08\textwidth]{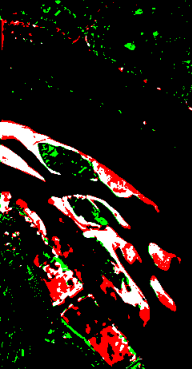}
}\hspace{-0.25cm}
\subfigure[]{
\includegraphics[width=0.08\textwidth]{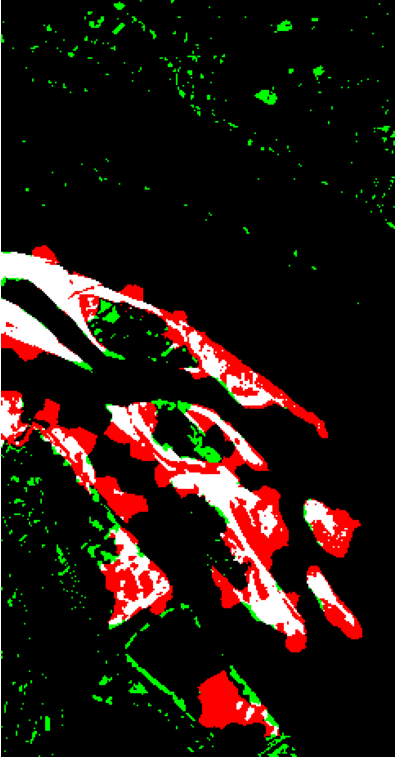}
}\hspace{-0.25cm}
\subfigure[]{
\includegraphics[width=0.08\textwidth]{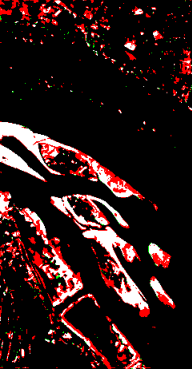}
}\hspace{-0.25cm}
\subfigure[]{
\includegraphics[width=0.08\textwidth]{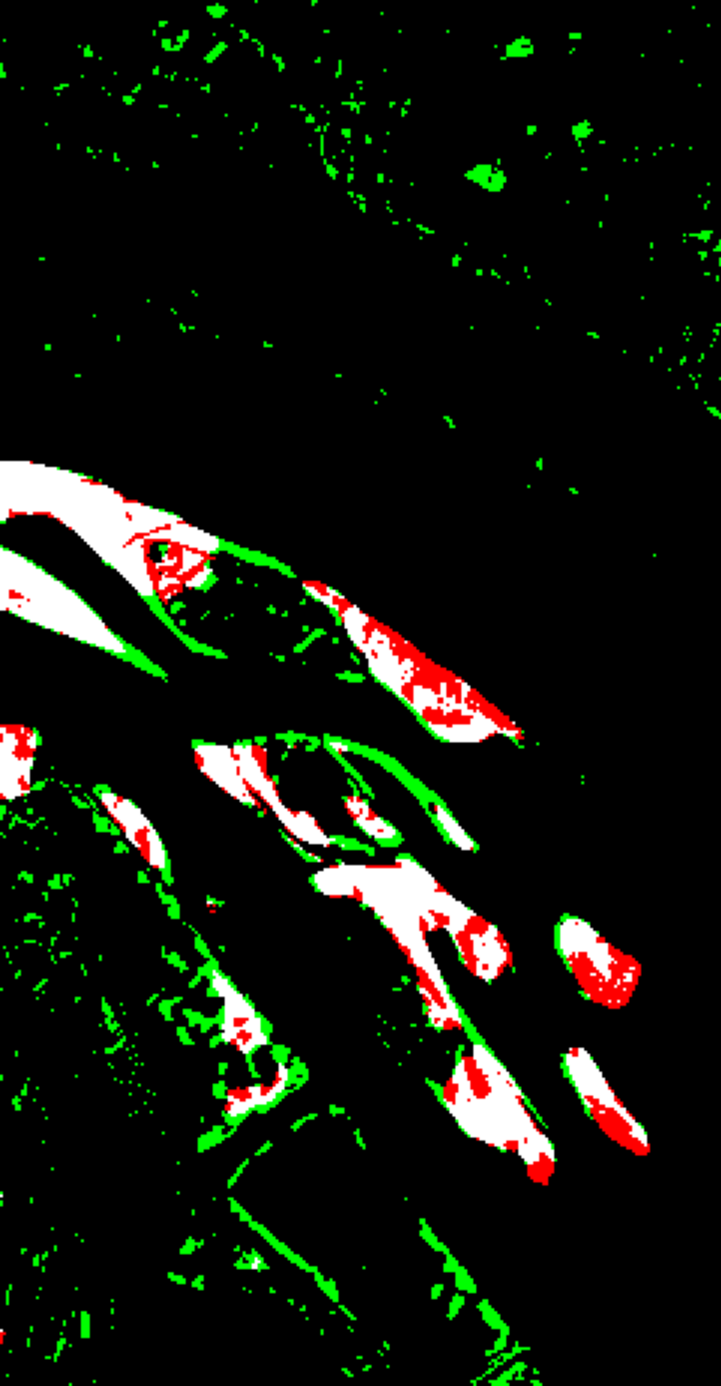}
}\hspace{-0.25cm}
\subfigure[]{
\includegraphics[width=0.08\textwidth]{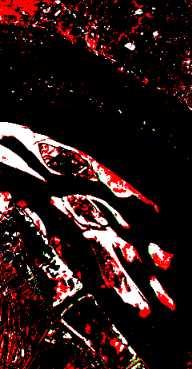}
}\hspace{-0.25cm}
\subfigure[]{
\includegraphics[width=0.08\textwidth]{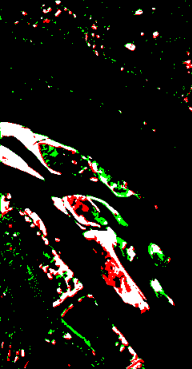}
}\hspace{-0.25cm}
\subfigure[]{
\includegraphics[width=0.08\textwidth]{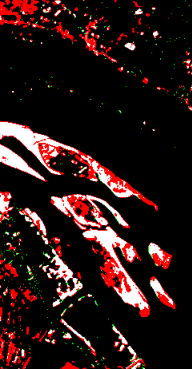}
}\hspace{-0.25cm}
\subfigure[]{
\includegraphics[width=0.08\textwidth]{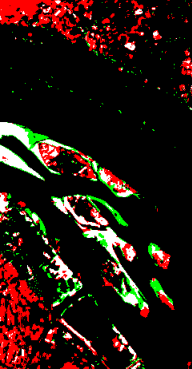}
}\hspace{-0.25cm}
\subfigure[]{
\includegraphics[width=0.08\textwidth]{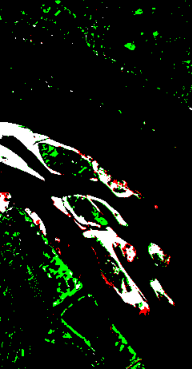}
}\hspace{-0.25cm}
\subfigure[]{
\includegraphics[width=0.081\textwidth]{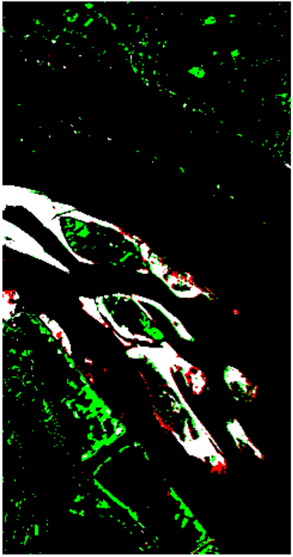}
}\hspace{-0.25cm}
\subfigure[]{
\includegraphics[width=0.08\textwidth]{gt_river}
}
\caption{Change detection results on the Jiangsu dataset, where FP and FN are marked in red and green, respectively. (a) HyperNet. (b) $\text{S}^3$Net. (c) QUEEN-$\mathcal{G}$. (d) ACFHAN. (e) CVA. (f) PCA-Kmeans. (g) TDRD. (h) PTCD. (i) HyperLUCID. (j) HyperLUCID (NVIDIA Orin Nano). (k) GT.
}
\label{fig:results_river}
\end{figure*}

\begin{figure*}[!t]
\centering
\subfigure[]{
\includegraphics[width=0.08\textwidth]{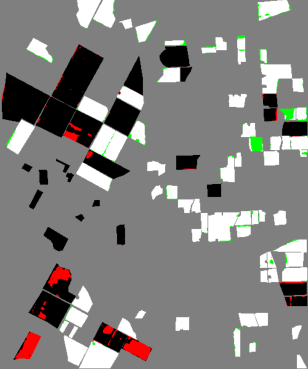}
}\hspace{-0.25cm}
\subfigure[]{
\includegraphics[width=0.08\textwidth]{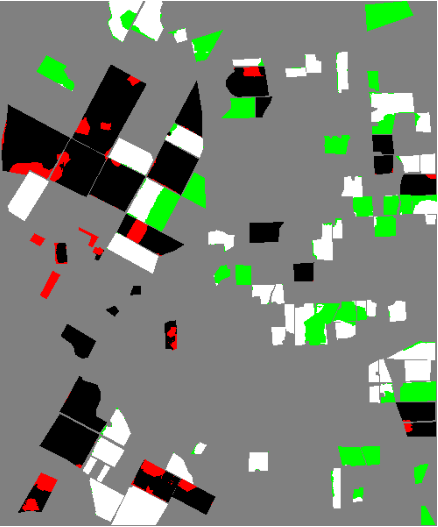}
}\hspace{-0.25cm}
\subfigure[]{
\includegraphics[width=0.08\textwidth]{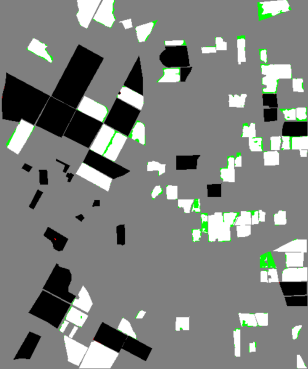}
}\hspace{-0.25cm}
\subfigure[]{
\includegraphics[width=0.08\textwidth]{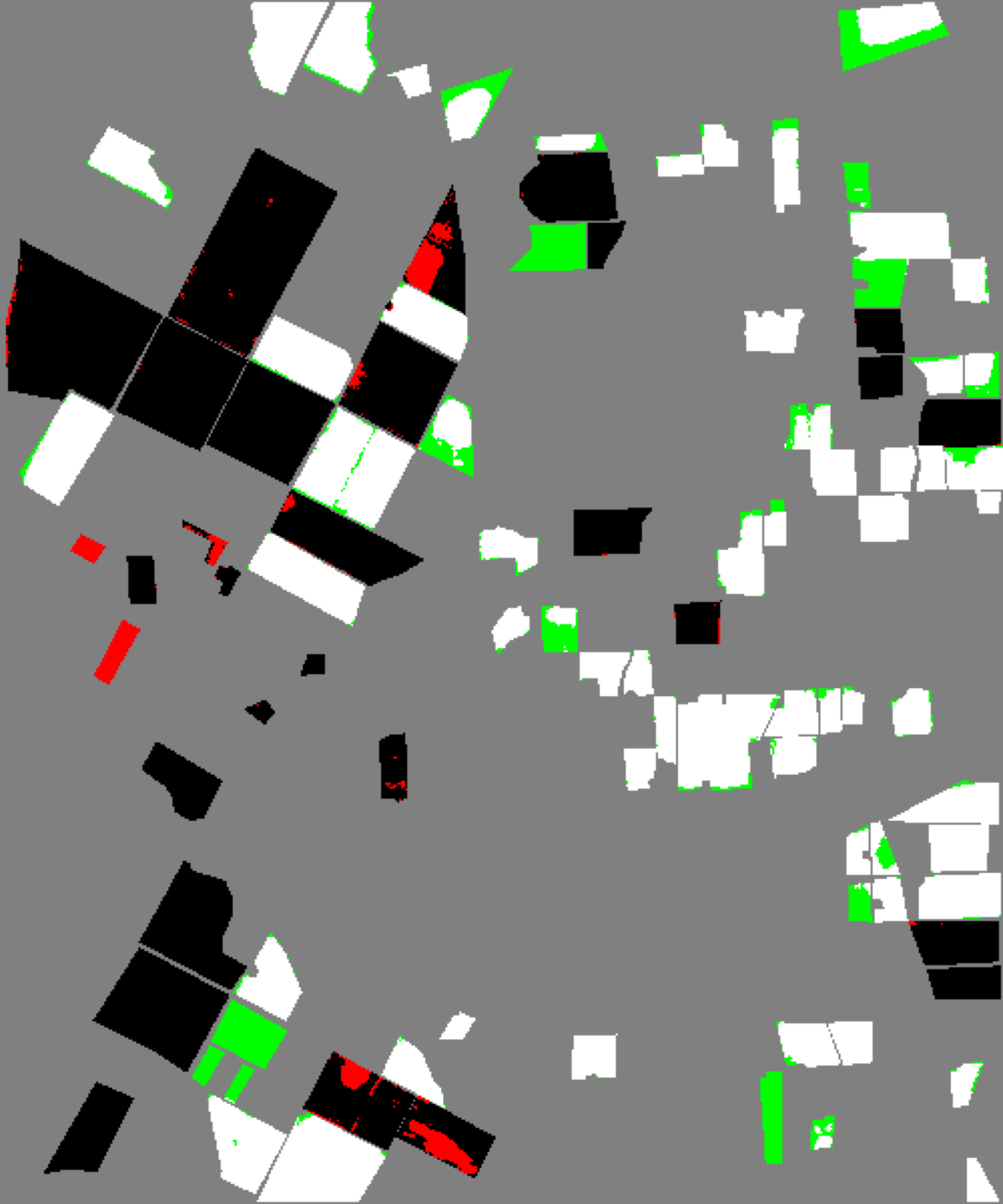}
}\hspace{-0.25cm}
\subfigure[]{
\includegraphics[width=0.08\textwidth]{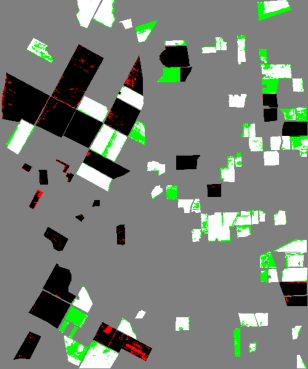}
}\hspace{-0.25cm}
\subfigure[]{
\includegraphics[width=0.08\textwidth]{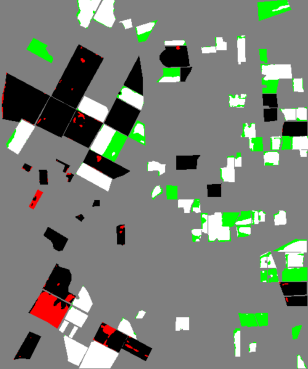}
}\hspace{-0.25cm}
\subfigure[]{
\includegraphics[width=0.08\textwidth]{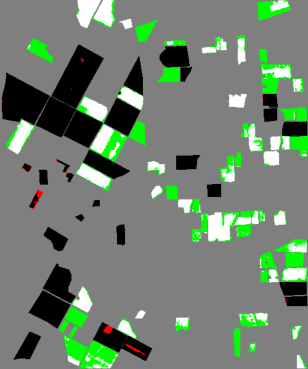}
}\hspace{-0.25cm}
\subfigure[]{
\includegraphics[width=0.08\textwidth]{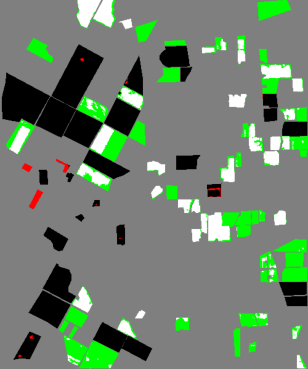}
}\hspace{-0.25cm}
\subfigure[]{
\includegraphics[width=0.08\textwidth]{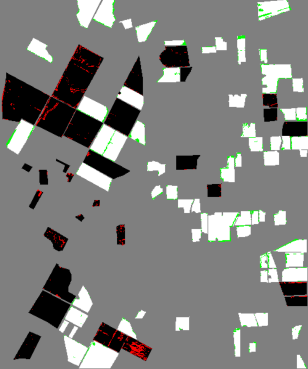}
}\hspace{-0.25cm}
\subfigure[]{
\includegraphics[width=0.0803\textwidth]{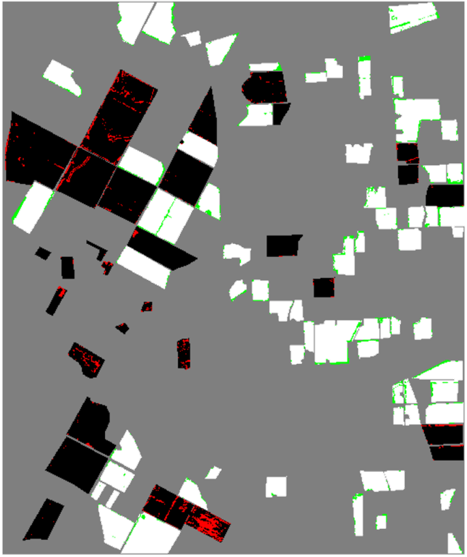}
}\hspace{-0.25cm}
\subfigure[]{
\includegraphics[width=0.08\textwidth]{gt_Bay}
}
\caption{Change detection results on the Bay Area dataset, where FP and FN are marked in red and green, respectively. (a) HyperNet. (b) $\text{S}^3$Net. (c) QUEEN-$\mathcal{G}$. (d) ACFHAN. (e) CVA. (f) PCA-Kmeans. (g) TDRD. (h) PTCD. (i) HyperLUCID. (j) HyperLUCID (NVIDIA Orin Nano). (k) GT.
}
\label{fig:results_Bay}
\vspace{-0.4cm}
\end{figure*}

\begin{figure*}[!t]
\centering
\subfigure[]{
\includegraphics[width=0.08\textwidth]{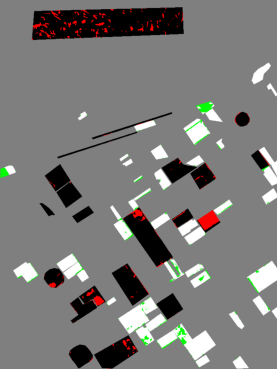}
}\hspace{-0.25cm}
\subfigure[]{
\includegraphics[width=0.08\textwidth]{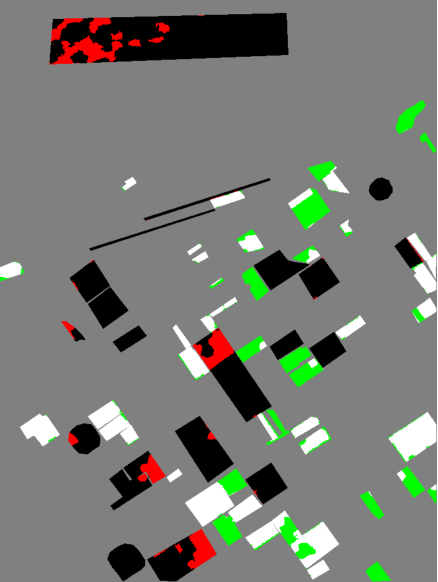}
}\hspace{-0.25cm}
\subfigure[]{
\includegraphics[width=0.08\textwidth]{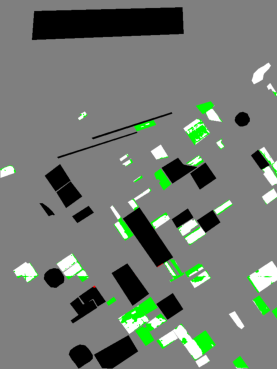}
}\hspace{-0.25cm}
\subfigure[]{
\includegraphics[width=0.08\textwidth]{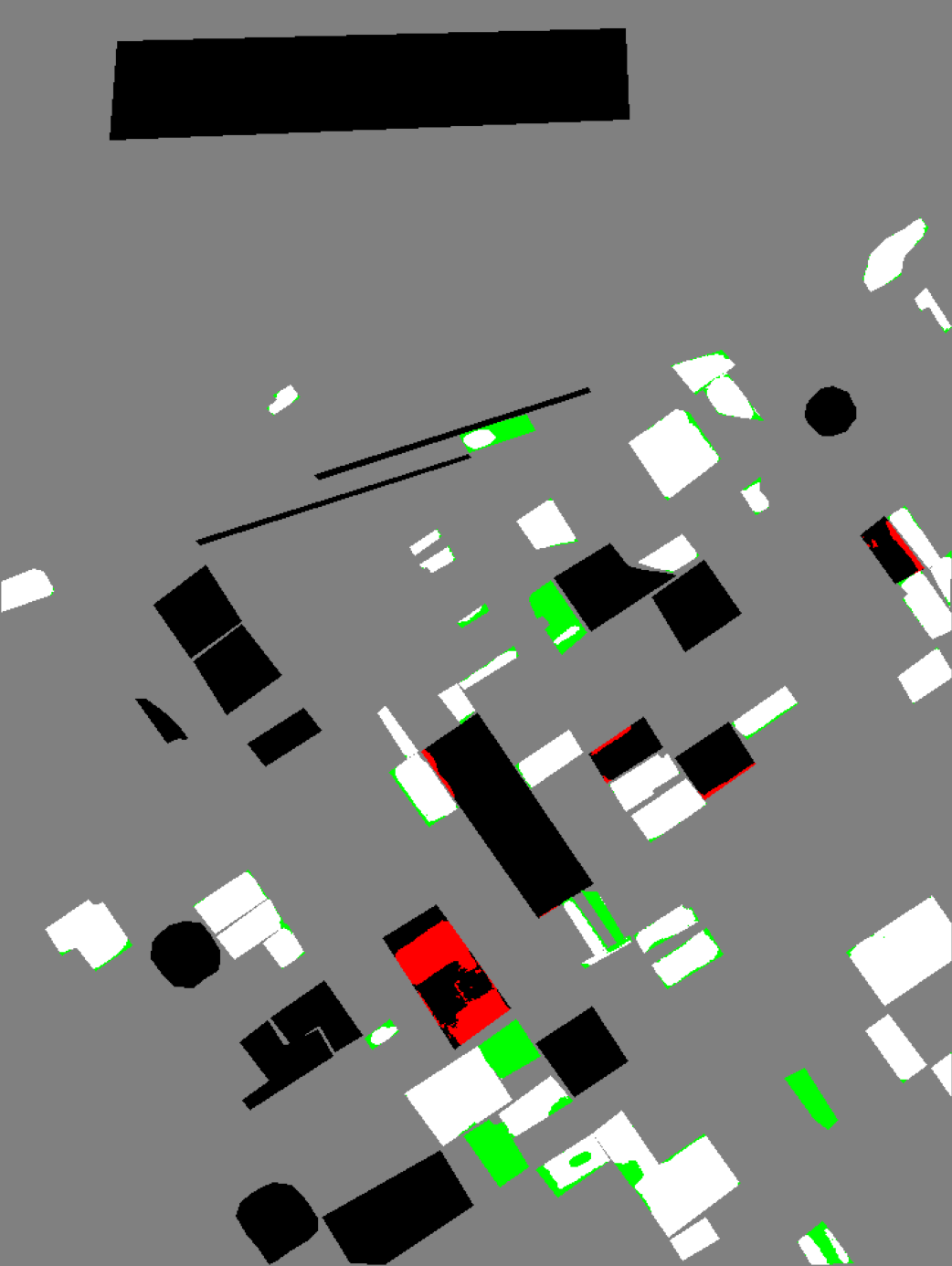}
}\hspace{-0.25cm}
\subfigure[]{
\includegraphics[width=0.08\textwidth]{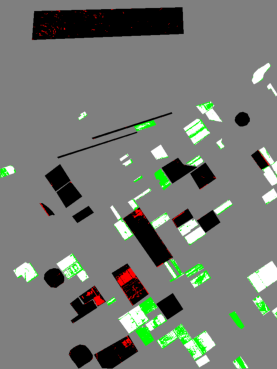}
}\hspace{-0.25cm}
\subfigure[]{
\includegraphics[width=0.08\textwidth]{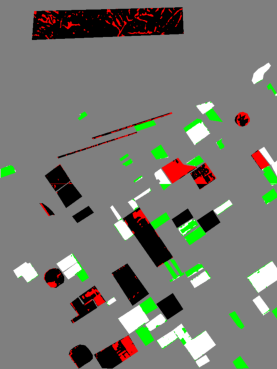}
}\hspace{-0.25cm}
\subfigure[]{
\includegraphics[width=0.08\textwidth]{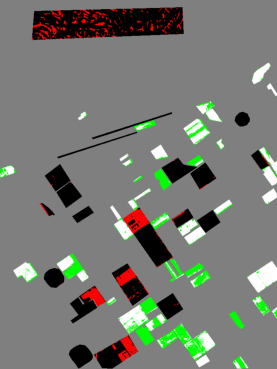}
}\hspace{-0.25cm}
\subfigure[]{
\includegraphics[width=0.08\textwidth]{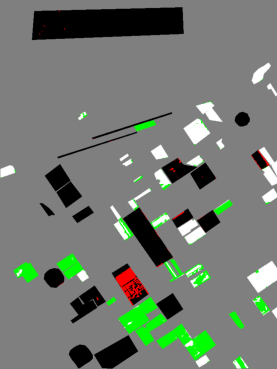}
}\hspace{-0.25cm}
\subfigure[]{
\includegraphics[width=0.08\textwidth]{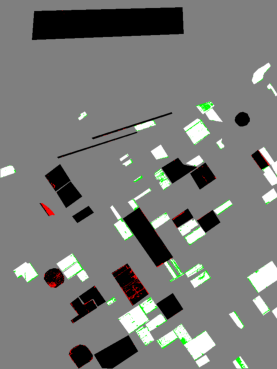}
}\hspace{-0.25cm}
\subfigure[]{
\includegraphics[width=0.0805\textwidth]{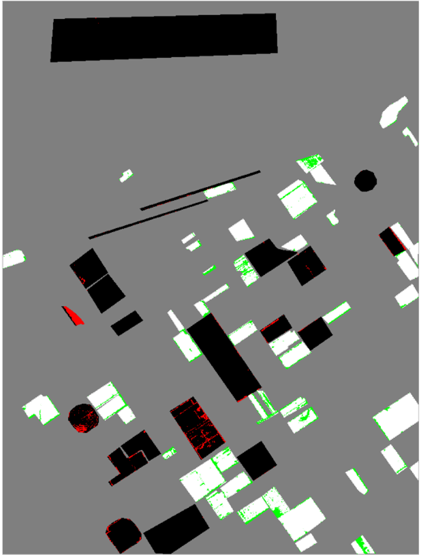}
}\hspace{-0.25cm}
\subfigure[]{
\includegraphics[width=0.08\textwidth]{gt_Barbara}
}
\caption{Change detection results on the Santa Barbara dataset, where FP and FN are marked in red and green, respectively. (a) HyperNet. (b) $\text{S}^3$Net. (c) QUEEN-$\mathcal{G}$. (d) ACFHAN. (e) CVA. (f) PCA-Kmeans. (g) TDRD. (h) PTCD. (i) HyperLUCID. (j) HyperLUCID (NVIDIA Orin Nano). (k) GT.
}
\label{fig:results_Barbara}
\end{figure*}

The quantitative comparisons on five real benchmark HCD datasets are summarized in Table \ref{tab:results}. Meanwhile, qualitative results are illustrated in Figures \ref{fig:results_Farm}, \ref{fig:results_Hermiston}, \ref{fig:results_river}, \ref{fig:results_Bay}, and \ref{fig:results_Barbara}, where FP and FN are highlighted in red and green, respectively.
Here, FP refers to pixels incorrectly detected as changed when they are unchanged, while FN denotes pixels that are mistakenly detected as unchanged when they are changed.

Across all datasets, HyperLUCID consistently demonstrates strong robustness under class imbalance, effectively preserving structural details and reducing both types of error. 
The safe sample selection strategy enhances robustness in complex scenarios, leading to reliable detection performance.
HyperLUCID also demonstrates excellent computational efficiency and consistently outperforms other deep learning-based methods in terms of runtime.
From the perspective of metric trade-offs, HyperLUCID emphasizes comprehensive change detection, with Pr slightly reduced and Re significantly improved.

In structured agricultural scenes such as Yancheng and Hermiston, where clear boundaries and regular patterns are dominant, existing methods such as HyperNet and $\text{S}^3$Net tend to produce excessive false positives, particularly along structural edges [cf. Figure~\ref{fig:results_Farm}(a), Figure~\ref{fig:results_Farm}(b), Figure~\ref{fig:results_Hermiston}(a), and Figure~\ref{fig:results_Hermiston}(b)].
In contrast, HyperLUCID yields more balanced predictions, effectively reducing both FPs and FNs while preserving boundary integrity, as evidenced by clearer boundaries and fewer misclassifications in Figure~\ref{fig:results_Farm}(i) and Figure~\ref{fig:results_Hermiston}(i).
Meanwhile, QUEEN-$\mathcal{G}$ achieves outstanding quantitative performance on these two datasets, which are consistent with its visual results, where relatively fewer FPs and FNs are observed [cf. Figure~\ref{fig:results_Farm}(c) and Figure~\ref{fig:results_Hermiston}(c)].

In river-dominant environments such as Jiangsu, where spectral signatures of water and surrounding regions are highly similar, existing methods such as HyperNet and $\text{S}^3$Net exhibit extensive false positives, particularly along high-contrast boundaries [cf. Figure~\ref{fig:results_river}(a) and Figure~\ref{fig:results_river}(b)].
HyperLUCID is more robust in this regard, achieving a higher precision of 0.871 while maintaining a superior F1 score of 0.734 (cf. Table~\ref{tab:results}), producing more accurate predictions in near-river regions [cf. Figure~\ref{fig:results_river}(i)].

In large-scale and subtle-change scenarios such as the Bay Area and Santa Barbara datasets, most methods fail to capture low-contrast changes, resulting in many false negatives (cf. Figure~\ref{fig:results_Bay} and Figure~\ref{fig:results_Barbara}).
Remarkably, HyperLUCID achieves strong performance, attaining (OA,~\!$\kappa$,~\!F1) values of (0.943,~\!0.885,~\!0.946) on Bay Area and (0.936,~\!0.865,~\!0.916) on Santa Barbara, respectively.
Qualitative results further show that HyperLUCID does produce more accurate predictions in subtle regions, particularly in the lower-right region of Bay Area and the bottom-right region of Santa Barbara [cf. Figure~\ref{fig:results_Bay}(i) and Figure~\ref{fig:results_Barbara}(i)].

\begin{figure}[t]
\begin{center}
\includegraphics[width=0.48\textwidth]{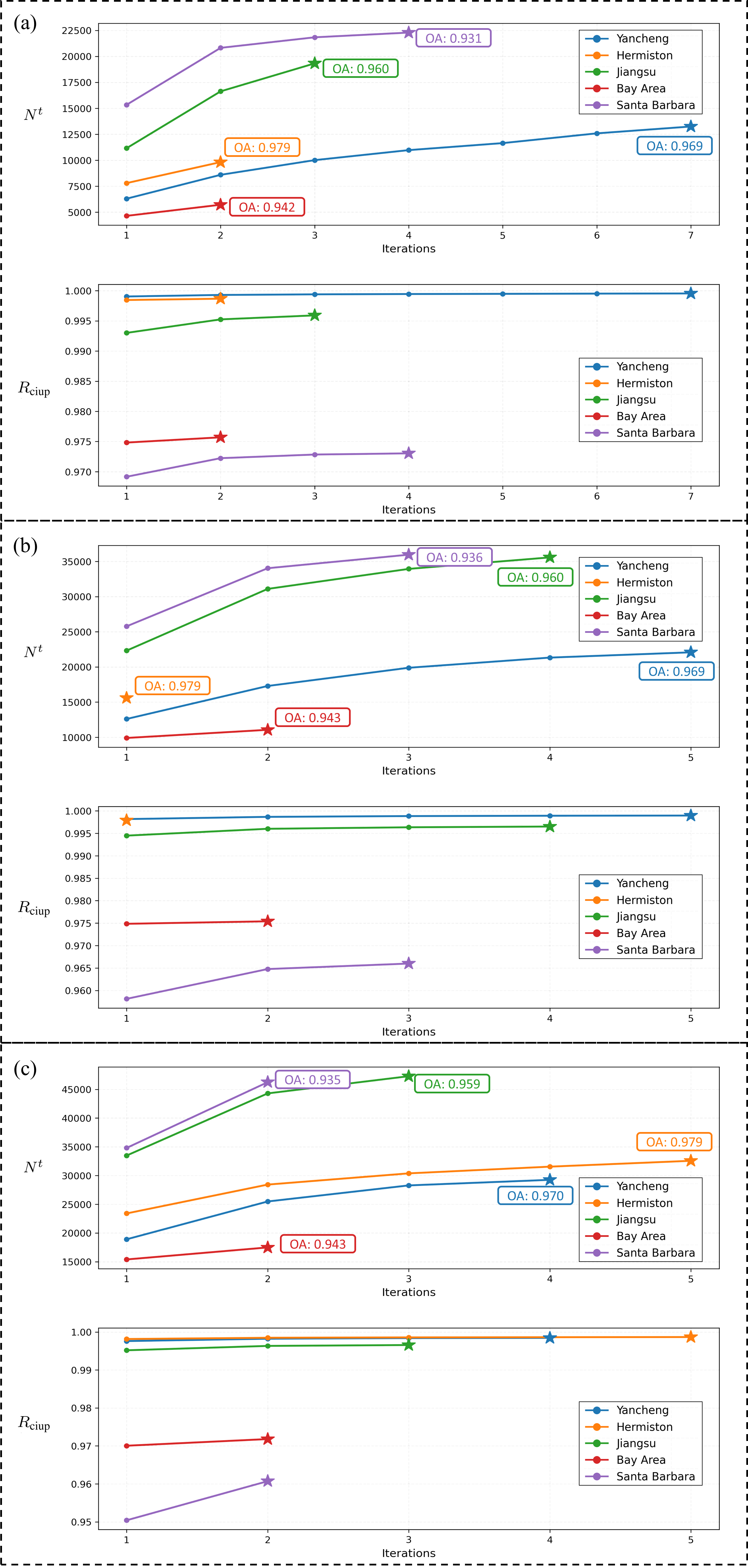}
%\vspace{0.05cm}
\caption{
For each benchmark testing data, we demonstrate the evolutions of $N^{t}$ and $R_\textrm{ciup}$ across the algorithmic iterations of HyperLUCID for (a) $R_\textrm{safe}= 10\%$, (b) $R_\textrm{safe}= 20\%$, and (c) $R_\textrm{safe}= 30\%$.
The iteration number marked by star means that the stopping criterion of HyperLUCID is met, thereby achieving the OA score (marked besides the star) that is not very sensitive to $R_\textrm{safe}$, thanks to the outlier-robustness brought by the L1-norm loss (cf. Property \ref{prop:L1robust}).
}\label{fig:fivecurves}
\end{center}
\vspace{-0.6cm}
\end{figure}

As discussed across diverse datasets and scene types, HyperLUCID stands out among unsupervised methods, delivering reliable performance in both quantitative metrics and qualitative analysis, particularly in terms of OA, $\kappa$, and F1.
In addition, the observed standard deviations are consistently low, indicating strong stability and robustness across multiple runs.
The success of the unsupervised HyperLUCID implies that the HyperCAD function $f$ has been effectively learned, alluding that the training set $\bm\Omega^\star$ is of high quality and has safely selected a very high ratio of unchanged samples, as will be substantiated by Figure \ref{fig:fivecurves}.
When considering runtime alongside performance, our method achieves the best balance between efficiency and effectiveness, benefiting from its lightweight network design.
% %
%
Furthermore, many peer methods show varying performance across datasets, particularly in large-scale and subtle-change scenarios (e.g., Bay Area and Santa Barbara), where they tend to achieve lower OA and $\kappa$.
Even strong methods such as QUEEN-$\mathcal{G}$ show degradation in certain cases with finer change patterns [cf. Figure~\ref{fig:results_Barbara}(c)], whereas HyperLUCID maintains more stable and consistent performance across diverse environments.
These results highlight the model’s strong generalization capability across diverse scenes and spectral characteristics. 
In particular, the proposed hyperspectral calibration function enables the model to better compensate for the variability of acquisition conditions, demonstrating excellent capability in detecting changes along complex boundaries and in subtle regions, where spectral variations are minimal.

As introduced in Section \ref{sec: introduction}, the representative self-supervised HyperSST \cite{A} works across the spectral-spatial-temporal domains.
According to the OA reported in \cite{A}, such a \textit{self-supervised} mechanism slightly outperforms our proposed \textit{unsupervised} scheme on the Yancheng data.
However, our unsupervised HyperLUCID clearly outperforms the self-supervised HyperSST in general, especially on the Hermiston data, for which HyperLUCID remarkably leads by more than 3\% OA (cf. Table \ref{tab:results} and \cite[Table I]{A}).
Moreover, for the very recent method ACFHAN \cite{ACFHAN}, its performance is good for the Yancheng/Hermiston/Jiangsu datasets, while its model tends to be conservative.
This can be seen from the much higher precision values (compared to the recall values; cf. Table \ref{tab:results}), and can also be observed from the considerable amount of the FN pixels (i.e., green pixels) in Figures \ref{fig:results_Farm}(d), 
\ref{fig:results_Hermiston}(d), 
\ref{fig:results_river}(d), 
\ref{fig:results_Bay}(d), and
\ref{fig:results_Barbara}(d).
In this regard, HyperLUCID is truly outstanding as both its FN and FP pixels are much fewer, especially for the sophisticated Bay Area and Santa Barbara  scenes [cf. Figures \ref{fig:results_Bay}(i), and
\ref{fig:results_Barbara}(i)].

\vspace{-0.4cm}
\subsection{Ablation Study}\label{sec:ablation}

\begin{table*}[t]
\centering
\caption{Ablation study on five real benchmark HCD datasets in terms of $\kappa$, F1, and running time (in seconds).
The boldfaced underlined number indicates the best performance, while the boldfaced number indicates the second-best.
Among the four models tested here, the first one is the original version of HyperLUCID; the second one is the HyperLUCID without the residual skip connection; the third one is the HyperLUCID trained using the naive Euclidean distance \eqref{eq:DI}; the fourth one is the LSTM model used to replace the CNN model in HyperLUCID.}
\renewcommand{\arraystretch}{1.3}
\begin{tabular}{|cccc|c|ccccc|}
\hline
\multicolumn{4}{|c|}{Modules}                                                                                                 & \multirow{2}{*}{\diagbox{Index}{Data}} & \multirow{2}{*}{Yancheng} & \multirow{2}{*}{Hermiston} & \multirow{2}{*}{Jiangsu} & \multirow{2}{*}{Bay Area} & \multirow{2}{*}{Santa Barbara} \\ \cline{1-4}
 Residual                     & SAM                          & CNN                           & LSTM                          &                                        &                           &                            &                          &                           &                                \\ \hline
\multirow{3}{*}{$\checkmark$} & \multirow{3}{*}{$\checkmark$} & \multirow{3}{*}{$\checkmark$} & \multirow{3}{*}{}             & $\kappa$($\uparrow$)                   & \textbf{0.926}            & \textbf{0.912}             & \textbf{0.713}           & \ubf{0.885}               & \ubf{0.865}                    \\
                              &                               &                               &                               & F1($\uparrow$)                         & \textbf{0.947}            & \textbf{0.923}             & \textbf{0.734}           & \ubf{0.946}               & \ubf{0.916}                    \\
                              &                               &                               &                               & Time($\downarrow$)                     & \textbf{3.576}            & \ubf{1.626}                & \textbf{4.162}           & \ubf{19.227}              & \ubf{50.626}                   \\ \hline
\multirow{3}{*}{}             & \multirow{3}{*}{$\checkmark$} & \multirow{3}{*}{$\checkmark$} & \multirow{3}{*}{}             & $\kappa$($\uparrow$)                   & \ubf{0.927}               & \ubf{0.913}                & \ubf{0.722}              & \ubf{0.885}               & \textbf{0.862}                 \\
                              &                               &                               &                               & F1($\uparrow$)                         & \ubf{0.948}               & \ubf{0.924}                & \ubf{0.742}              & \ubf{0.946}               & \textbf{0.914}                 \\
                              &                               &                               &                               & Time($\downarrow$)                     & 5.638                     & 3.673                      & 8.311                    & 29.066                    & 88.596                         \\ \hline
\multirow{3}{*}{$\checkmark$} & \multirow{3}{*}{}             & \multirow{3}{*}{$\checkmark$} & \multirow{3}{*}{}             & $\kappa$($\uparrow$)                   & 0.880                     & 0.902                      & 0.695                    & 0.530                     & 0.677                          \\
                              &                               &                               &                               & F1($\uparrow$)                         & 0.913                     & 0.915                      & 0.727                    & 0.712                     & 0.787                          \\
                              &                               &                               &                               & Time($\downarrow$)                     & \ubf{3.037}               & \textbf{3.173}             & \ubf{1.995}              & \textbf{20.603}           & \textbf{64.851}                \\ \hline
\multirow{3}{*}{$\checkmark$} & \multirow{3}{*}{$\checkmark$} & \multirow{3}{*}{}             & \multirow{3}{*}{$\checkmark$} & $\kappa$($\uparrow$)                   & 0.921                     & 0.747                      & 0.689                    & \textbf{0.807}                     & 0.837                          \\
                              &                               &                               &                               & F1($\uparrow$)                         & 0.944                     & 0.792                      & 0.711                    & \textbf{0.922}                     & 0.900                          \\
                              &                               &                               &                               & Time($\downarrow$)                     & 15.709                    & 18.887                     & 29.232                   & 108.408                   & 303.444                        \\ \hline
\end{tabular}
\label{tab:ablation_1}
\end{table*}

In this section, we demonstrate that the residual strategy does significantly accelerate the HyperLUCID algorithm.
The effectiveness of residual learning stems from its ability to model only the subtle differences between the acquired and calibrated spectra, making the optimization process more efficient.
To verify the benefits of the residual strategy, we remove the skip connection from the network $f$ (cf. Figure \ref{fig:netf}) and evaluate its performance.
The results are presented in the first and second rows of Table~\ref{tab:ablation_1}, showing that the convergence speed does decline significantly when the residual strategy is not applied.
Although the model without residuals achieves comparable quantitative performance on all datasets, it suffers from a substantial increase in running time; for example, it takes more than twice as long on the Hermiston and Jiangsu datasets compared to the model with residual connections.
In summary, using residual connections speeds up computation significantly while maintaining satisfactory results.

We further show that the choice of the SAM distance \eqref{eq: SAM} is critical in the HyperLUCID model.
Specifically, we compare the naive Euclidean distance function defined in \eqref{eq:DI} with the SAM distance.
Illumination variation is a key causal factor to misclassify unchanged pixels as changed ones, while SAM is a robust pseudo-label selection criterion as discussed below \eqref{eq: SAM}.
As shown in the first and third rows of Table~\ref{tab:ablation_1}, the SAM module does outperform the Euclidean distance function in all quantitative metrics.
However, in the Yancheng and Jiangsu datasets, adopting the Euclidean distance function achieved faster convergence.
This phenomenon can be attributed to the simplistic nature of the Euclidean distance, which limits the model's capacity to further identify safe samples and therefore leads to earlier convergence (and weaker performance).
These results validate the effectiveness of SAM-based safe sample selection in improving the quality of pseudo-labels and enhancing overall detection accuracy.

Finally, we demonstrate that the proposed simple yet effective network architecture (cf. Figure \ref{fig:netf}) is sufficient for the calibration task.
Notably, although the gated mechanisms adopted in the long short-term memory (LSTM) \cite{LSTM} (for effectively preserving important signals over long sequences while discarding noise) looks suitable for modeling the long spectrum information of HSIs, experiments show that increasing the model complexity by employing more sophisticated mechanisms would not lead to better HCD performance.
As shown in the first and last rows of Table~\ref{tab:ablation_1}, replacing the CNN module with the LSTM module results in inferior performance across all datasets, both in terms of quantitative metrics and running time.
In particular, adopting the LSTM module instead of CNN results in a running time more than five times longer across all datasets.
In summary, increasing the complexity of the model not only degrades overall performance but also results in significantly slower convergence.

\begin{figure}[t]
\begin{center}
\includegraphics[width=0.48\textwidth]{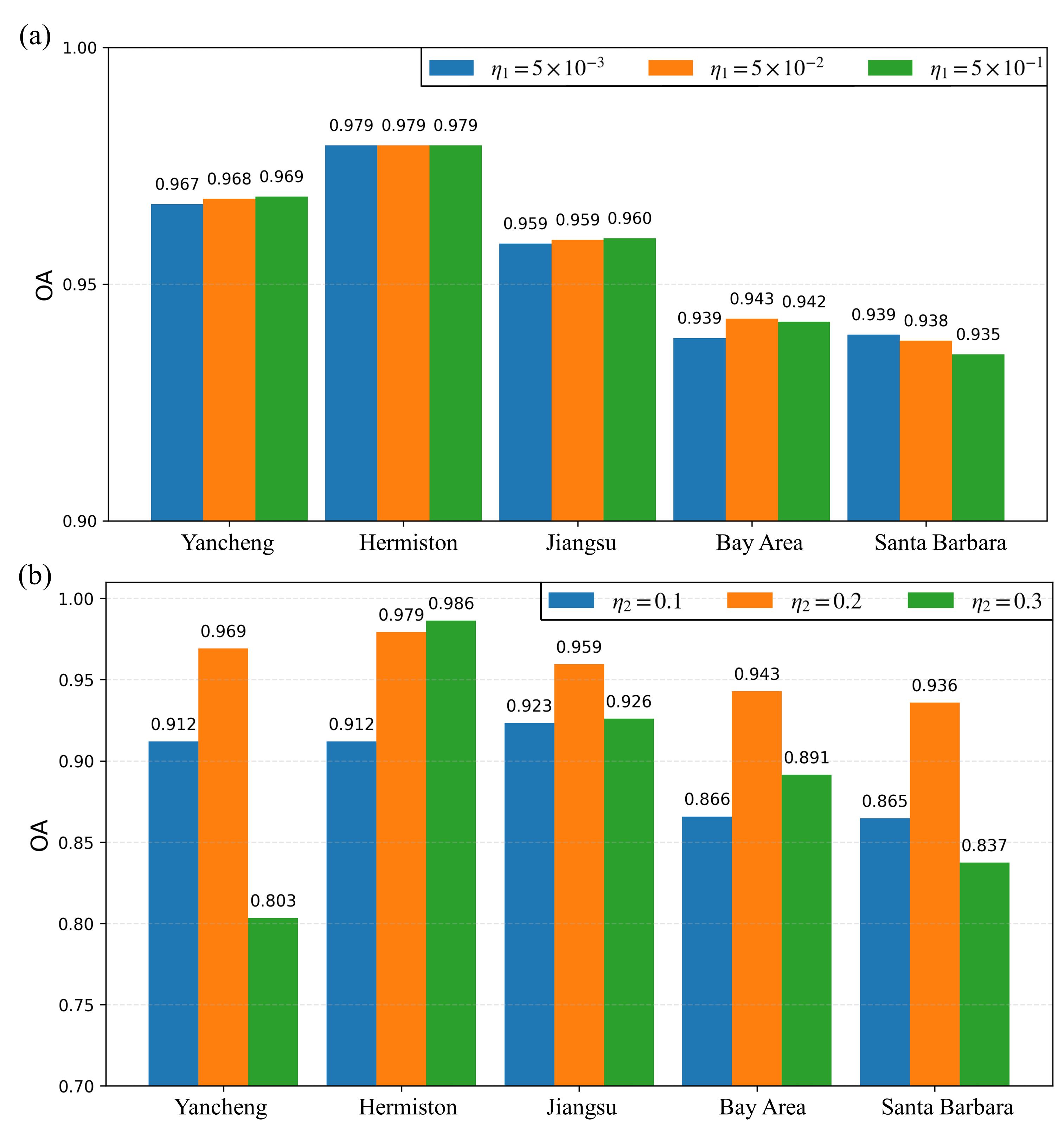}
\caption{ 
Demonstration of the local optimality of the unified hyperparameter setting of $(\eta_1,\eta_2)$ across various scenarios, including (a) the threshold standard deviation $\eta_1:=0.05$, and (b) the threshold SAM value $\eta_2:=0.2$.
The performance is not sensitive to $\eta_1$, but somewhat sensitive to $\eta_2$.
Fortunately, the setting of $(\eta_1,\eta_2):=(0.05,0.2)$ yields generally strong HCD performances across various benchmark testing datasets.
}\label{fig:SAM_stoppingrule}
\end{center}
\vspace{-0.5cm}
\end{figure}

\vspace{-0.5cm}
\subsection{Discussion}\label{sec:discussion}

To have a user-friendly algorithm, we begin by discussing how to unify the setting of the hyperparameters in HyperLUCID.
To have a convincing default setting, besides the most widely tested benchmark HCD images (i.e., Yancheng data, Hermiston data, and Jiangsu data) \cite{QUEENG}, we also consider additional testing images (i.e., Bay Area data, and Santa Barbara data), aiming at identifying a unified hyperparameter setting across all these benchmark testing scenarios.
Though this is challenging, some theoretical thoughts or empirical validations are presented below for identifying the unified setting of the ratio of safe samples $R_\textrm{safe}:=20\%$, the threshold standard deviation $\eta_1:=0.05$, and the threshold SAM value $\eta_2:=0.2$.

Recall the definition that $N^{t}\triangleq|\bm\Omega^{t}|$ denote the number of estimated unchanged pixels in $\bm\Omega^{t}$ at iteration $t$, and accordingly define the ratio of correctly identified unchanged pixels in $\bm\Omega^{t}$ as $R_\textrm{ciup}\triangleq \frac{|\bm\Omega^{t}\cap\bU|}{|\bm\Omega^{t}|}$.
In Figure \ref{fig:fivecurves}, we consider $R_\textrm{safe}\in\{10\%, 20\%, 30\%\}$, where for each testing data we demonstrate the evolutions of $N^{t}$ and $R_\textrm{ciup}$ across the algorithmic iterations of HyperLUCID.
Before discussing the hyperparameters, we remark that $R_\textrm{ciup}$ is remarkably greater than $99\%$ for Yancheng/Hermiston/Jiangsu data, greater than $97\%$ for Bay Area data, and greater than $96\%$ for Santa Barbara data, indicating that the key condition (A) set in Section \ref{sec:algo} is indeed almost satisfied.
Note that it is impossible to require the ideal case of $R_\textrm{ciup}=100\%$ in the fully unsupervised framework (as no ground-truth labeling is available), but HyperLUCID has an embedded robustness mechanism against the violation of condition (A) (as discussed in Property \ref{prop:L1robust}).
Also, early misclassifications do not seem to reinforce bias in subsequent
iterations, even in the heterogeneous regions (e.g., Yancheng/Jiangsu data, and Bay Area data), as demonstrated in Figure \ref{fig:fivecurves}.
For example, in Figure \ref{fig:fivecurves}(b), although there are misclassified pixels in $\bm\Omega^t$ (i.e., $R_\textrm{ciup}\neq 1$) from the very beginning for the Jiangsu data, $R_\textrm{ciup}$ remains high across all iterations, alluding that early misclassification does not deteriorate the quality of incremental training.
This should be attributed to the outlier-robust learning scheme (cf. Property \ref{prop:L1robust}).

Let us discuss the hyperparameters.
First, from Figure \ref{fig:fivecurves}, if we increase the ratio of safe samples $R_\textrm{safe}$ from $20\%$ to $30\%$, the number of outliers $N^{t}(1-R_\textrm{ciup})$ (i.e., the number of pixels in $\bm\Omega^{t}$ but not in $\bU$) significantly increases by more than 200 for the Bay Area data.
In other words, $R_\textrm{safe}:=30\%$ introduces significantly more outliers, and is hence not regarded as a safe ratio to sift unchanged pixels.
By contrast, a smaller ratio of $R_\textrm{safe}:=10\%$ will collect the safe samples slowly.
For example, by decreasing $R_\textrm{safe}$ from $20\%$ to $10\%$, the number of iterations increases from 3 to 4 for the Santa Barbara data, but the slower convergence even decreases the OA from 93.6\% to 93.1\%, because the number of correctly identified safe samples (i.e., $N^{t} R_\textrm{ciup}$) significantly decreases by around 13000, hampering the effective learning of HyperLUCID.
Thus, we empirically set $R_\textrm{safe}:=20\%$, which yields generally strong HCD performances across diverse scenarios, as proved in Table \ref{tab:results}.
Second, regarding the stopping criterion, the threshold standard deviation $\eta_1:=0.05$ is actually a commonly seen setting.
As can be seen from Figure \ref{fig:SAM_stoppingrule}(a), setting a smaller $\eta_1$ (e.g., 0.005) could cause significantly inferior performance (e.g., the Hermiston  data), while setting a larger $\eta_1$ (e.g., 0.5) generally yields similar or slightly weaker performances.
Third, as for the threshold SAM value $\eta_2:=0.2$ for deciding the final detection map $\bC^\star$, some of the best hyperspectral analysis algorithms could achieve SAM values of around 0.2 to 2 degrees \cite[Table III]{HyperCSI}.
Thus, $\eta_2:=0.2$ is a safe setting to argue that a pixel is unchanged; both a smaller setting of $\eta_2:=0.1$ or a larger setting of $\eta_2:=0.3$ lead to obvious performance degradation as can be seen from Figure \ref{fig:SAM_stoppingrule}(b).
Therefore, the setting of $(\eta_1,\eta_2):=(0.05,0.2)$ does empirically yield strong HCD performances across various datasets.

We remark that some unsupervised clustering methods may also be adapted for the bi-temporal HCD task.
For example, we tried to modify the single-temporal hyperspectral clustering algorithm, called diffusion-based spatial-spectral image reconstruction and clustering (DSIRC) \cite{cui2022unsupervised}, by manually aligning the clustering labels between $\bX$ and $\bY$ and thresholding the clustering-based spectral SAM differences to derive the change map, where the best-performing threshold is empirically found to be 0.18.
The modified DSIRC achieves 0.942 OA, 0.863 kappa, 0.857 precision, 0.958 recall, and 0.905 F1 score on the Yancheng dataset.

Finally, HyperLUCID has remarkably achieved reductions of several orders of magnitude in FLOPs and over one order of magnitude in model parameters compared to existing methods (cf. Figure \ref{fig:flopsradius}), substantially lowering both computational and memory demands for the onboard computing environments.
We further corroborate these advantages through deployment on a resource-constrained embedded platform, the NVIDIA Jetson Orin Nano \cite{scalcon2024ai}. 
Experimental results [cf. Figures \ref{fig:results_Farm}(j), 
\ref{fig:results_Hermiston}(j), 
\ref{fig:results_river}(j), 
\ref{fig:results_Bay}(j), and
\ref{fig:results_Barbara}(j)] show that the performance of HyperLUCID remains nearly identical to that obtained on the desktop system, indicating that HyperLUCID does not rely on excessive computational resources to maintain its effectiveness. 
This consistency across platforms confirms its practical deployability, making HyperLUCID a strong candidate for real-world onboard systems.

\vspace{-0.2cm}
\section{Conclusion and Future Work}\label{sec: conclusion}

We have proposed the HyperLUCID algorithm (i.e., Algorithm \ref{algo:HyperLUCID}) for change detection.
It is designed based on the idea of compensating the variability of acquisition conditions (cf. Figure \ref{fig:HyperCAD}).
We formally establish the idea as an image calibration problem in Section \ref{sec:problem}, and then implement the idea using a fully unsupervised iterative mechanism in Section \ref{sec:algo}.
To facilitate the computational efficiency, the calibration function is cast as a lightweight group convolution network (cf. Figure \ref{fig:netf}).
Employing the inherent resemblance between the acquired spectra and the calibrated spectra, the calibration network is judiciously designed using a residual learning architecture, which further speeds up the HyperLUCID algorithm.
Experiments demonstrate that a more sophisticated network architecture could hamper the convergence speed or even degrade the detection performance.
The distance function measuring the differences between hyperspectral pixels is also critical for the calibration task, and experiments show that the adopted SAM distance \eqref{eq: SAM} is less sensitive to the acquisition condition variability.
The proposed HyperLUCID algorithm (graphically illustrated in Figure \ref{fig:flow}) is suitable for onboard edge-computing deployment, and has achieved state-of-the-art HCD results on several real benchmark datasets under a unified parameter/hyperparameter setting.
Our unsupervised algorithm (even outperforming semi-supervised methods) and the ideas developed in this paper can be used to solve many real-world remote sensing applications (e.g., satellite-driven automatic detection of unregistered illegal buildings) and can even be extended to develop RGB/radar/medical imagery change detection algorithms in the future.

Another future research line would be to further address the potential violation of the assumption that low-SAM pixels are mostly unchanged; this limitation is currently addressed only by Property \ref{prop:L1robust}.
Although HyperLUCID works well on benchmark HCD datasets, some extreme real-world situations are not explicitly addressed yet, including drastic seasonal changes and natural disasters (often inducing large changed-area ratios).
It is also valuable to include an automatic mechanism to address severe misregistration especially when $(\bX,\bY)$ are acquired by different satellites/sensors.
Moreover, although ablation study shows that HyperLUCID does not rely on sophisticated networks, the users would have an option to plug a more advanced calibration network (e.g., quantum deep network, QUEEN \cite{QUEENG}) into the HyperLUCID framework, if severe nonlinear spectral variability or other complicated effects cannot be ignored.
Furthermore, HyperLUCID is more sensitive to $\eta_2$, whose robust default value has been empirically identified as $\eta_2:=0.2$.
To mitigate the sensitivity, a future work would be to theoretically derive (e.g., from hyperspectral geometry \cite{zhuang2019regularization}) a scene-adaptive optimal setting of $\eta_2$ in an automatic manner.
Finally, software optimization frameworks such as TensorRT, pruning, quantization, and layer fusion \cite{edgedeploy2} can be adopted for edge-device profiling of HyperLUCID in the future.

\vspace{-0.2cm}
\appendix

\subsection{Raw Edge-Device Profiling on NVIDIA Jetson Orin Nano}\label{sec:proof Lemma 1}

This appendix simply demonstrates the feasibility of edge-device implementation based on NVIDIA Jetson Orin Nano, without using any software optimization tricks (e.g., pruning, quantization, and layer fusion), which are out of the scope of this paper.
Remarkably, even if we directly profile HyperLUCID on Jetson Orin Nano without any software optimization, the peak memory usage is still far below the maximum allowable limit of Jetson Orin Nano (8GB).
For example, the peak memory usage for Yancheng data is only 3.34\% of the maximum allowable (cf. Table \ref{tab:jetson_edge_profiling_power}).

It is worth noting that deploying an algorithm from a desktop/server environment to an edge board often introduces additional runtime and resource constraints due to limited computation, memory, and power resources on edge devices \cite{edgedeploy1}.
Prior studies have shown that deep learning inference on edge devices involves non-negligible runtime overheads, while software optimization frameworks such as TensorRT, pruning, quantization, and layer fusion can substantially improve embedded-GPU inference efficiency \cite{edgedeploy2}.
For example, edge-oriented object-detection systems often require additional optimization techniques, such as knowledge distillation, TensorRT-based deployment, or hardware-aware optimization, to improve execution efficiency on resource-constrained devices \cite{edgedeploy3}.
We left these implementation tricks for HyperLUCID as future works.

Since the main contribution of this work is not edge acceleration or TensorRT-level optimization, we report the raw GPU-ready profiling results of the complete HyperLUCID pipeline, rather than attempting to close the performance gap between a desktop GPU workstation and an edge device through additional system-level optimization.
The results are summarized in Table \ref{tab:jetson_edge_profiling_power}.

Since HyperLUCID is an iterative calibration and pseudo-label updating framework, rather than a pure feed-forward inference model, we report the execution time instead of conventional latency.
Specifically, following NVIDIA's TensorRT benchmarking practice\footnote{\url{https://docs.nvidia.com/deeplearning/tensorrt/latest/performance/benchmarking.html}}, one-time warm-up and setup overheads are excluded from the measured execution window.
The reported time measures the complete HyperLUCID processing stage from GPU-ready input tensors to final change-map generation.

In addition to execution time, Table \ref{tab:jetson_edge_profiling_power} also reports peak GPU memory usage, throughput, average board power, and estimated energy consumption.
Throughput is measured in MPixels/sec. (i.e., million pixels processed per second), and energy is estimated as the product of average board power and GPU-ready execution time.
The relatively longer execution time on the Santa Barbara dataset is mainly due to its very larger spatial size.
Even though, the peak memory usage is only 39.04\% (i.e., 3123.125MB/8GB) of the
maximum allowable of Jetson Orin Nano (cf. Table \ref{tab:jetson_edge_profiling_power}), even if we do not adopt any software optimization trick.
These results verify the practical deployability of HyperLUCID on resource-constrained edge hardware.

\begin{table}[t] \centering \caption{Raw edge-device profiling of HyperLUCID on NVIDIA Jetson Orin Nano. 
Execution time (Exec. Time), peak memory usage (Peak Mem.), throughput, average power (Avg. Power), and Energy are measured in seconds (sec.), MB, MPixels/sec., Watt (W), and Joule (J), respectively, as reported in this table. 
% Energy is estimated as average power multiplied by execution time.
}\label{tab:jetson_edge_profiling_power}\renewcommand{\arraystretch}{1.12} \setlength{\tabcolsep}{3.6pt} \scriptsize 
\resizebox{\linewidth}{!}{
\begin{tabular}{c|c|c|c|c|c}\hline Dataset  & Exec. Time  & Peak Mem. & Throughput & Avg. Power  & Energy  \\ \hline Yancheng &  13.422 & 273.402 & 0.0047 & 11.290 & 151.537 \\  \hline Hermiston &  4.945 & 304.508 & 0.0158 & 12.546 & 62.036 \\ \hline Jiangsu  & 11.885 & 347.376 & 0.0094 & 13.426 & 159.562  \\ \hline Bay Area  & 33.338 & 1287.393 & 0.0022 & 14.132 & 471.144  \\ \hline Santa Barbara  & 221.435 & 3123.125 & 0.0006 & 8.687 & 1923.585 \\ \hline\end{tabular} 
}
\vspace{-0.2cm}
\end{table}

\bibliography{ref}
\begin{IEEEbiography}[{\resizebox{0.9in}{!}{\includegraphics[width=1in,height=1.25in,clip,keepaspectratio]{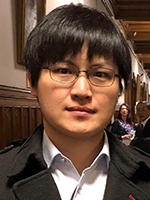}}}]
{\bf Chia-Hsiang Lin}
(S'10-M'18-SM'24)
received the B.S. degree in electrical engineering and the Ph.D. degree in communications engineering from National Tsing Hua University (NTHU), Taiwan, in 2010 and 2016, respectively.
From 2015 to 2016, he was a Visiting Student of Virginia Tech,
Arlington, VA, USA.

He is currently a Professor with the Department of Electrical Engineering, National Cheng Kung University (NCKU), Taiwan, and serves as the Technical Director of the Industrial Technology Research Institute (ITRI), Taiwan.
Before joining NCKU, he held research positions with The Chinese University of Hong Kong, HK (2014 and 2017),
NTHU (2016-2017),
and the University of Lisbon (ULisboa), Lisbon, Portugal (2017-2018).
He was an Assistant Professor with the Center for Space and Remote Sensing Research, National Central University, Taiwan, in 2018, and a Visiting Professor with ULisboa, in 2019.
His research interests include network science,
quantum computing,
convex geometry and optimization, blind signal processing, and imaging science.

Dr. Lin received the Emerging Young Scholar Award (The 2030 Cross-Generation Program) from National Science and Technology Council (NSTC), from 2023 to 2027,
the Future Technology Award from NSTC, in 2022,
the Outstanding Youth Electrical Engineer Award from The Chinese Institute of Electrical Engineering (CIEE), in 2022,
the Best Young Professional Member Award from IEEE Tainan Section, in 2021,
the Prize Paper Award from IEEE Geoscience and Remote Sensing Society (GRS-S), in 2020, and The 3rd Place from AIM Real World Super-Resolution Challenge at IEEE International Conference on Computer Vision (ICCV), in 2019.
He received the Ministry of Science and Technology (MOST) Young Scholar Fellowship, together with the EINSTEIN Grant Award, from 2018 to 2023.
In 2016, he was a recipient of the Outstanding Doctoral Dissertation Award from the Chinese Image Processing and Pattern Recognition Society and the Best Doctoral Dissertation Award from the IEEE GRS-S.
\end{IEEEbiography}

\begin{IEEEbiography}[{\resizebox{0.9in}{!}{\includegraphics[width=1in,height=1.25in,clip,keepaspectratio]{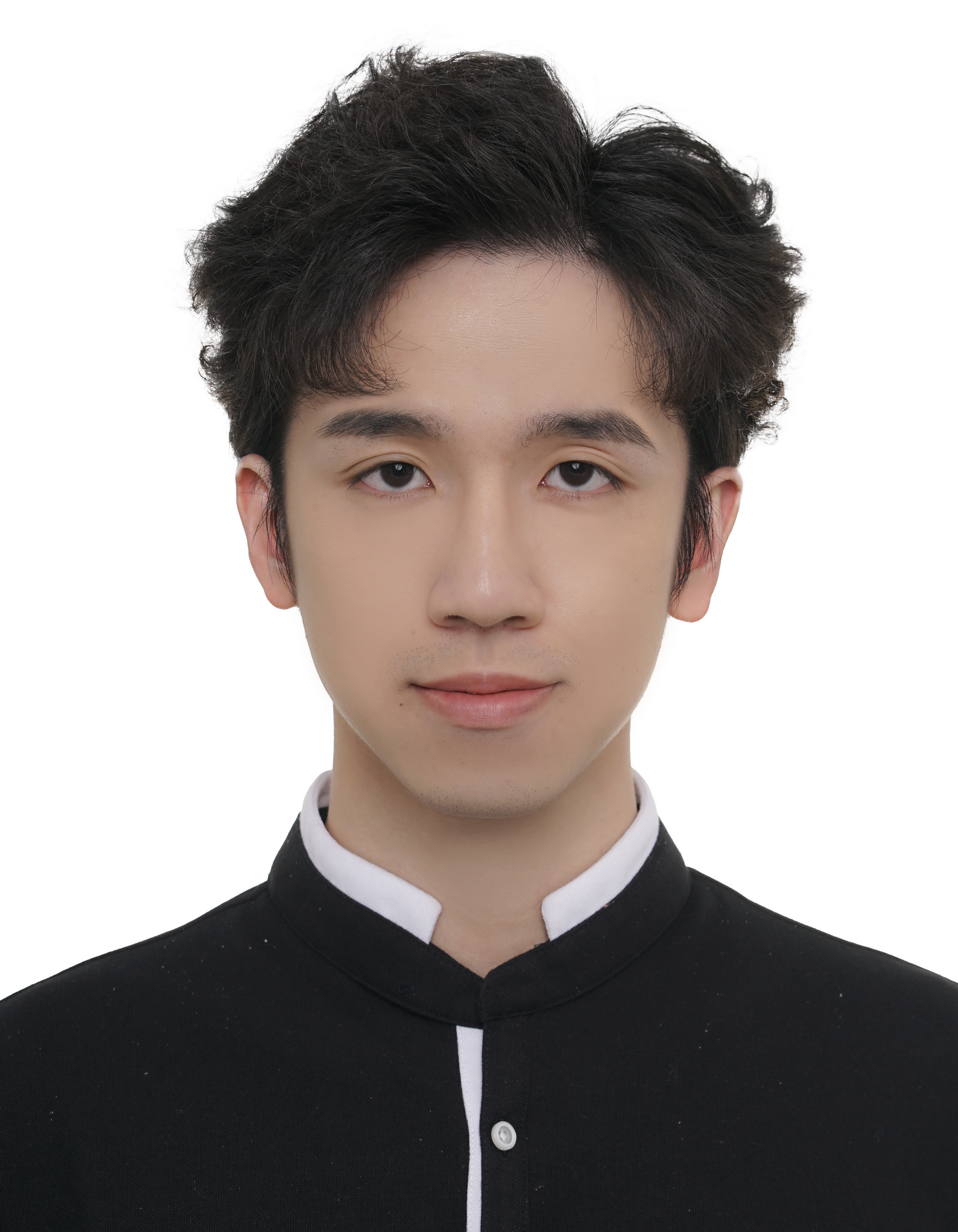}}}]
	{\bf Shih-Min Hsu}
	(S'24) is currently a Ph.D. student with the Intelligent Hyperspectral Computing Laboratory (IHCL), Department of Electrical Engineering, National Cheng Kung University (NCKU), Tainan, Taiwan.
    His research interests include deep learning, bioinformatics and biomedical imaging, hyperspectral remote sensing, and quantum deep learning.
    He recently received the prestigious Ph.D. Student Scholarship Funding Award for the 2025-2026 term from the Ministry of Education, Taiwan.
\end{IEEEbiography}

\begin{IEEEbiography}[{\resizebox{0.9in}{!}{\includegraphics[width=1in,height=1.25in,clip,keepaspectratio]{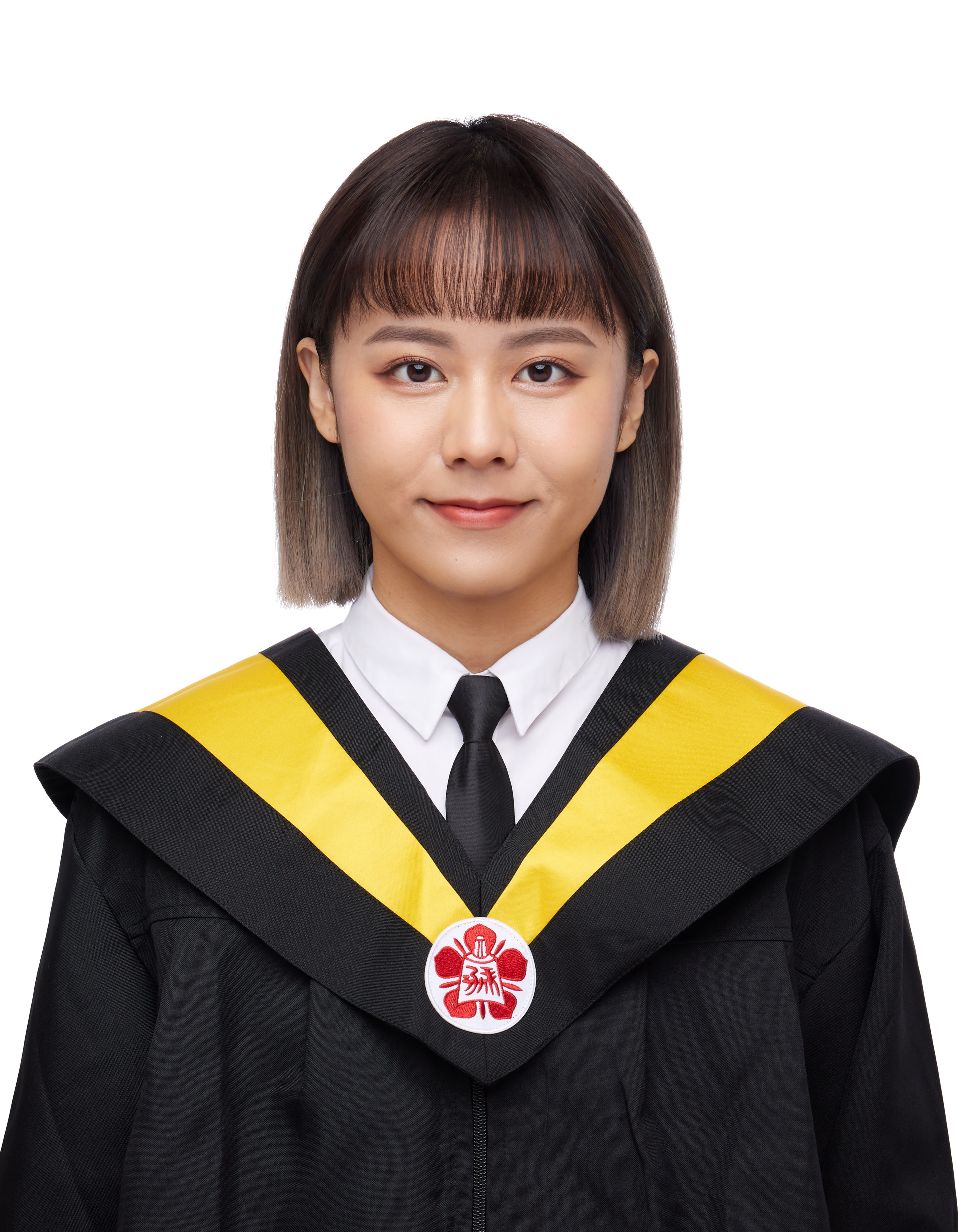}}}]
{\bf Ching-Yun Liang}
received the B.S. degree (in 3.5 years) from the Department of Physics, National Cheng Kung University, Taiwan, in 2024.
She is currently pursuing the M.S. degree with the Intelligent Hyperspectral Computing Laboratory (IHCL), Department of Electrical Engineering, National Cheng Kung University (NCKU), Tainan, Taiwan. 

Her research interests include deep learning, quantum computing, convex optimization, unsupervised learning, and hyperspectral remote sensing.
\end{IEEEbiography}

\begin{IEEEbiography}[{\resizebox{1in}{!}{\includegraphics[width=1in,height=1.25in,clip,keepaspectratio]{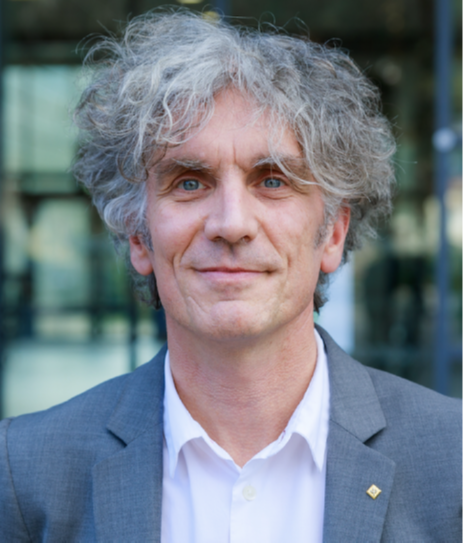}}}]
    {\bf Jocelyn Chanussot}
    (IEEE Fellow)
    received the M.Sc. degree in electrical engineering from the Grenoble Institute of Technology (Grenoble INP), Grenoble, France, in 1995, and the Ph.D. degree from the Université de Savoie, Annecy, France, in 1998.

    From 1999 to 2023, he has been with Grenoble INP, where he was a Professor of signal and image processing.
    He is currently a Research Director with INRIA, Grenoble.
    His research interests include image analysis, hyperspectral remote sensing, data fusion, machine learning, and artificial intelligence.
    He has been a visiting scholar at Stanford University (USA), KTH (Sweden), and NUS (Singapore).
    Since 2013, he is an Adjunct Professor of the University of Iceland.
    In 2015-2017, he was a visiting professor at the University of California, Los Angeles (UCLA).
    He holds the AXA chair in remote sensing and is an Adjunct Professor at the Chinese Academy of Sciences, Aerospace Information Research Institute, Beijing, China.
    
    Dr. Chanussot is the founding President of IEEE Geoscience and Remote Sensing French chapter (2007-2010), which received the 2010 IEEE GRSS Chapter Excellence Award.
    He was the Vice-President of the IEEE Geoscience and Remote Sensing Society, in charge of meetings and symposia (2017-2019).
    He is an Associate Editor for the IEEE Transactions on Geoscience and Remote Sensing, the IEEE Transactions on Image Processing, and the Proceedings of the IEEE.
    He was the Editor-in-Chief of the IEEE Journal of Selected Topics in Applied Earth Observations and Remote Sensing (2011-2015).
    In 2014 he served as a Guest Editor for the IEEE Signal Processing Magazine.
    He is a Fellow of the IEEE, an ELLIS Fellow, a Fellow of AAIA, a member of the Institut Universitaire de France (2012-2017), and a Highly Cited Researcher (Clarivate Analytics/Thomson Reuters, since 2018).
\end{IEEEbiography}

\begin{IEEEbiography}[{\resizebox{1in}{!}{\includegraphics[width=1in,height=1.25in,clip,keepaspectratio]{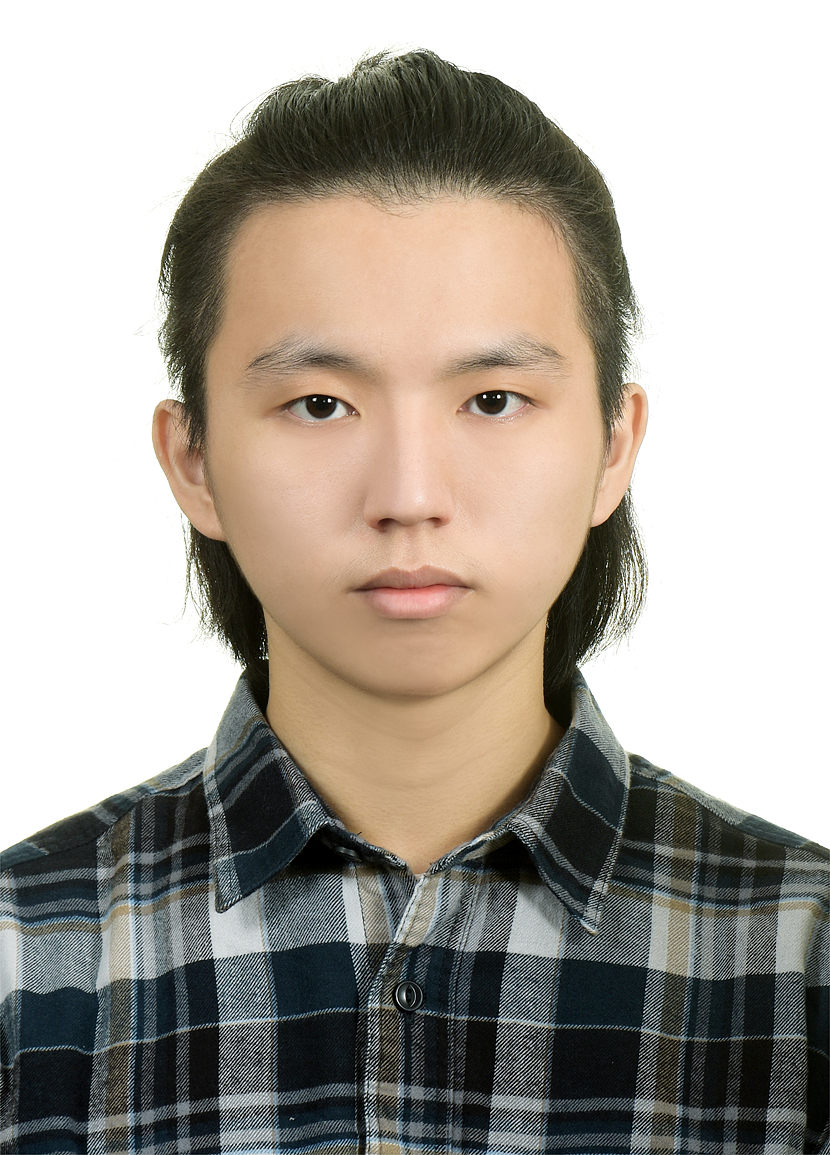}}}]
{\bf Jhih-Yan Chen}
received the B.S. degree from the Department of Electronics Engineering from Chang Gung University, Taiwan, in 2023.
He was pursuing the M.S. degree with the Intelligent Hyperspectral Computing Laboratory (IHCL), Department of Electrical Engineering, National Cheng Kung University (NCKU), Tainan, Taiwan.
His research interests include small-data deep learning, hyperspectral change detection, and convex optimization.
He just received the Honorary Membership of The Phi Tau Phi Scholastic Honor Society in 2025.
\end{IEEEbiography}

\end{document}